\documentclass[aps,prd,twocolumn,superscriptaddress,nofootinbib]{revtex4-1}

\usepackage{aas_macros}
\usepackage[utf8]{inputenc}
\usepackage[T1]{fontenc}
\usepackage[english]{babel}
\usepackage{mathrsfs}
\usepackage{amsmath}
\usepackage{amssymb}
\usepackage{tipa}
\usepackage{ntheorem}
\usepackage{braket}
\usepackage{graphicx}
\usepackage{tabularx}

\usepackage{bm}
\usepackage{wasysym}
\usepackage{enumitem}
\usepackage{stmaryrd}
\usepackage[breaklinks=true,colorlinks,citecolor=blue,linkcolor=blue,urlcolor=blue]{hyperref}
\usepackage{tikz}
\usepackage{footmisc}
\usetikzlibrary[patterns]
\usetikzlibrary{shapes}

\usepackage[varg]{txfonts}

\usepackage{mathtools}

\usepackage{placeins}
\usepackage{multirow}

\makeatletter
\newskip\@bigflushglue \@bigflushglue = -100pt plus 1fil

\def\bigcentering{\let\\\@centercr\rightskip\@bigflushglue
\leftskip\@bigflushglue
\parindent\z@\parfillskip\z@skip}

\makeatother

\allowdisplaybreaks[1]

\newcommand{\ghost}[1]{}

\newcommand{\R}{\mathbb R}
\newcommand{\C}{\mathbb C}

\newcommand{\dd}{\mathrm{d}}
\newcommand{\deriv}[2]{\frac{\dd #1}{\dd #2}}

\renewcommand{\sec}[1]{\left\langle {#1} \right\rangle}
\newcommand{\secderiv}[2]{\sec{ \deriv{#1}{#2} }}

\renewcommand{\vec}[1]{\boldsymbol {#1}}
\newcommand{\uvec}[1]{\boldsymbol {\hat #1}}
\newcommand{\basis}[1]{\boldsymbol {\hat e}_{\mathbf #1}}

\newcommand{\Real}{\mathfrak{Re}}
\newcommand{\Imag}{\mathfrak{Im}}

\newcommand{\im}{\mathrm{i}}
\newcommand{\e}{\mathrm{e}}

\newcommand{\UnitSphere}{\mathcal{S}}

\newcommand{\norm}[1]{\lvert #1 \rvert}

\DeclareMathOperator{\arctantwo}{arctan2}

\begin{document}

   \title{Secular dipole-dipole stability of magnetic binaries}

    \author{C. Aykroyd}
    \email{christopher.aykroyd@obspm.fr}
    \affiliation{SYRTE, Observatoire de Paris, Université PSL, CNRS, Sorbonne Université, LNE, 61 avenue de l'Observatoire, 75014 Paris, France}
    
    \author{A. Bourgoin}
    \affiliation{SYRTE, Observatoire de Paris, Université PSL, CNRS, Sorbonne Université, LNE, 61 avenue de l'Observatoire, 75014 Paris, France}
    \affiliation{Universit\'e Paris-Saclay, Universit\'e Paris Cit\'e, CEA, CNRS, AIM, F-91191, Gif-sur-Yvette, France}
       
    \author{C. Le Poncin-Lafitte}
    \affiliation{SYRTE, Observatoire de Paris, Université PSL, CNRS, Sorbonne Université, LNE, 61 avenue de l'Observatoire, 75014 Paris, France}
       
    \author{S.~Mathis}
    \affiliation{Universit\'e Paris-Saclay, Universit\'e Paris Cit\'e, CEA, CNRS, AIM, F-91191, Gif-sur-Yvette, France}

    \author{M.-C. Angonin}
    \affiliation{SYRTE, Observatoire de Paris, Université PSL, CNRS, Sorbonne Université, LNE, 61 avenue de l'Observatoire, 75014 Paris, France}
    
   % \date{}

% \abstract{}{}{}{}{} 
% 5 {} token are mandatory

\begin{abstract}
  The presence of strong large-scale stable magnetic fields in a significant portion of early-type stars, white dwarfs, and neutron stars is well-established. Despite this, the origins of these fields remain a subject of ongoing investigation, with theories including fossil fields, mergers, and shear-driven dynamos. One potential key for understanding the formation of these fields could lie in the connection between magnetism and binarity. Indeed, magnetism can play a significant role in the long-term orbital and precessional dynamics of binary systems. In gravitational wave astronomy, the advanced sensitivity of upcoming interferometric detectors such as LISA and Einstein Telescope will enable the characterization of the orbital inspirals of compact systems, including their magnetic properties.
  A comprehensive understanding of the dynamics of magnetism in these systems is necessary for the interpretation of the gravitational wave signals and to avoid biases in the calibration of instruments. This knowledge can additionally be used to create new magnetic population models and provide insight into the nature and origins of their internal magnetic fields.
  % }
  % {
  The aim of this study is to investigate the secular spin precession dynamics of binary systems under pure magnetic dipole-dipole interactions, with a focus on stars with strong, stable, and predominantly dipolar fields.
  % }
  % {
  We employ an orbit-averaging procedure to the spin precession equations from which we derive an effective secular description. By minimizing the magnetic interaction energy of the system, we obtain the configurations of spin equilibrium and their respective stabilities. Finally, we also derive a set of conditions required for the validity of our assumptions to hold.
  % }
  % {
  We show that among the four states of equilibrium, there is a single secular state that is globally stable, corresponding to the configuration where the spin and magnetic axes of one star are reversed with respect to the companions', and orthogonal to the orbital plane. Our results are compared to traditional methods of finding instantaneous states of equilibrium, in which orbital motion is generally neglected. Finally, we provide analytical solutions in the neighbourhood of the stable configuration, that can be used to derive secular orbital evolution in the context of gravitational wave astronomy.
\end{abstract}
  % {}

   % \keywords{magnetic fields -- celestial mechanics -- binaries: close -- stars: neutron -- white dwarfs -- stars: early type } %  
   
%
%________________________________________________________________

\maketitle

\section{Introduction}

Close to $10\%$ of early-type massive main sequence (MMS) stars host stable large-scale magnetic fields, ranging from $3 \times 10^2$ to $3 \times 10^4$~G \citep{Grunhut2013, Ferrario_2015a, Shultz2019}. Meanwhile, it is estimated that $20$--$25\%$ of the white dwarf (WD) population is magnetic \citep{Bagnulo_2021}, with detected fields between $10^3$ and $10^9$~G. For neutron stars (NS), surface field strengths are gauged on the order of $10^{8}$--$10^{13}$~G in classical radio pulsars \citep{Phinney2003}, and reach up to $10^{14}$--$10^{15}$~G in magnetars \citep{Kaspi2017}.
%, and $10\%$ of neutron stars (NS) are magnetars, found in the high-field range between $10^{14}$ and $10^{15}\,\text{G}$ \AB{A citation would be nice here}. 
%\todo{cite, single WD (13\% host high-fields > $10^5$G)}
The origins of magnetic fields are highly debated for both main-sequence stars and for compact objects; we present below the prevailing theories for each respective class of stars.

In main-sequence late-type stars,
it is considered established that external magnetic fields are driven by dynamo action in the outer convective zone \citep{Brun2017,Brun2022}. Conversely, this channel is less likely to be the case for hot, massive stars ($M > 1.5 M_\odot$), which maintain a radiative envelope and inner convective core: any dynamo-based explanation must resolve the challenge of transporting the magnetic field towards the outer surface faster than the stellar evolution timescale \citep{Charbonneau_2001}. 
In fact, in radiative stars, magnetic fields are believed to decay in diffusivity timescales estimated longer than their host's main-sequence lifetime \citep{Moss_2003}.
Early direct evidence via Zeeman spectropolarimetry has long since shown that chemically peculiar Ap and Bp stars, which represent around 10\% of early-type A/B stars \citep{Ferrario_2015a}, host strong secularly stable magnetic fields, with strengths uncorrelated with stellar rotation --- as should be the case for dynamo-fed fields --- and geometry largely captured by oblique dipole rotor models \citep[see e.g{.}][]{Stibbs1950, Borra_1982}).
These fields have therefore been suggested to have `fossil' origin, remnants of a prior stellar evolution stage and effectively frozen into the plasma \citep{Moss_1987, Mathis2011}. The exact field formation process is a topic of debate, but multiple plausible mechanisms have been proposed --- ranging from accumulated magnetic flux captured from the interstellar cloud at birth, to protostar mergers and pre-main-sequence dynamos \citep{Ferrario_2009}. % also independent of mass, luminosity and rotation.
More recent studies such as the `B Fields in OB Stars' (BOB; \citep{Morel2014}) and the `Magnetism in Massive Stars' (MiMeS; \citep{Wade_2016_MiMeS, Shultz2019}) surveys also confirmed compatible magnetic incidence and properties on more general O/B-type massive stars.
Numerical and semi-analytical magneto-hydrodynamic computations have established the existence of long-term-stable internal field configurations consistent with non-convective bodies such as radiative stars, NS and WDs, which favours the fossil field scenario \citep{Braithwaite_2004, Braithwaite_2006, Braithwaite_2008, %Braithwaite_2009,
Duez_2010, Fuller2023}. These fields configurations are composed of both toroidal and poloidal components that stabilise each other \citep{Tayler1980, Braithwaite_2009, Akgun2013}; outside the star, the toroidal field is attenuated and mainly the poloidal component is visible. 
%Purely poloidal or purely toroidal configurations are long since known to develop instabilities (Tayler), 
These results were found to reproduce the general characteristics of observations: the roughly off-center dipolar structure, the independence from stellar spin, and finally, the strong field amplitude \citep{Braithwaite_2004, Duez2011}. % Axissymetric & non-symmetric
Nevertheless, the fossil field hypothesis is not without its challenges. For example, only a small fraction of MMS stars host observable fields, with a sharp dearth of weak-field objects; the precise mechanism for field formation, stability and evolution must explain this cutoff. 
It has been suggested that there are thresholds to field strength below which shear or convection instabilities develop \citep[see e.g{.}][]{Auriere_2007, Gaurat2015, Jouve2020, Jermyn2021}, or that, in some stars, the time needed to reach an equilibrium becomes longer than the age in the main sequence, due to the Coriolis force produced by rapid rotation \citep{Braithwaite2013}. An additional challenge to fossil fields is the existence of a great scarcity of magnetism in close binaries, as low as 2\% incidence \citep{Carrier_2002, Alecian_2014}. In fact, there is a single known doubly-magnetic close binary up to date, the $\epsilon$-Lupi system \citep{Pablo_2019}. If the fossil scenario is indeed to be the main field formation channel, it is plausible to expect a similar magnetic incidence in binaries and in single stars.
However, \cite{Vidal_2017} suggests that tidal instabilities in binary pairs can disrupt the magnetic fields via turbulent Joule diffusion within a few millions years, potentially explaining the scarcity of strong-field binaries. Alternatively, it has been argued that interstellar clouds with strong magnetic fields are harder to fragment \citep{Commencon_2011}, yielding selection biases towards less magnetic binary systems. Other alternatives have also been suggested to address this challenge, such as the merger scenarios \citep{Ferrario_2009, Ferrario_2015a, Schneider_2016, Schneider2019}. In these scenarios, coalescing main sequence stars and/or protostars would generate strong enough shear to drive dynamo action, yielding a single magnetic byproduct star. Such hypotheses are in line with the prediction that around 8\% of MMS stars originate from mergers \citep{de_Mink_2014}, and naturally explain the lack of magnetic binaries. Nevertheless, at this stage, no channel can be completely favoured over another.% On the other hand, merger theories lack ... \todo{6\% blue stragglers}.

In the compact object community there is a somewhat analogous debate. On one side, classical fossil theories defend that magnetic white dwarfs (MWDs) and NS are derived from Ap/B and O-type stars respectively, and that their fields must persist from the main sequence or red giant phase \citep{Tout_2004, Ferrario2005, Wickramashinghe_2005}. % or post-MS dynamo?
Another possibility lifted by \cite{Stello_2016, Bagnulo_2021} is that internal dynamos in the convective cores of intermediate-mass and massive stars --- externally invisible during the main sequence phase --- might develop into strong stable fields by flux compression as the stellar core collapses into a WD. These fields would then be slowly revealed as the WD sheds its outer layers, and decay in secular Ohmic timescales.
On the other side of the debate, merger theories \citep{Tout_2008, Ferrario_2020} advocate an intimate link between magnetism and binarity; in such scenarios, two common-envelop stars would generate a magnetic field through differential rotation and merge to form a strong-field MWD. Closely interacting systems which failed to completely merge might instead develop into magnetic cataclysmic variables. Finally, alternative theories support the operating of some dynamo mechanism during the cooling of the WD --- e.g.\ during the crystallisation convection of the core in a rapidly rotating WD \citep{Isern_2017, Schreiber_2021}.
In support of the fossil hypothesis in WDs and NS is the striking similarity in magnetic flux between this group and the main sequence A/B/O stars, as well as the long field decay timescales --- estimated to be on the order of tens or hundreds of billions of years \citep{Cumming_2002, Ferrario_2015_MWDs}. Conversely, the progenitors of non-magnetic WDs would be low-mass stars, which are known to harbor relatively weak dynamo-driven fields. This contrasting spectral origin is consistent with the observation that MWDs are on average more massive than their non-magnetic counterparts \citep{Liebert_1988, Bagnulo_2021}. However, merger scenarios would also naturally explain such disparity.
Against the fossil field hypothesis, it has been argued that there is an insufficient volume density of Ap/Bp stars to by itself account for the high occurance of MWDs, as required for the classical fossil theory to hold \citep{Kawka2004, Ferrario2005, Bagnulo_2021}. Furthermore, many surveys have pointed out a sparsity of known detached binaries composed of a MWD plus a non-degenerate companion \citep{Liebert_2005, Ferrario_2015_MWDs}, whereas conversely, magnetism amongst cataclysmic variables is ubiquitous, with about a quarter of these WDs reaching the high-field range ($B \geq 1\,\text{MG}$).
This has propelled the suggestion that magnetism and binarity in WD systems are intrinsically connected, leading to the advent of merger hypotheses. However, as pointed out by \cite{Landstreet_2020, Bagnulo_2021}, %common search methods are hampered by strong biases, and 
at the present moment, at least five such detached MWD binary systems are known, and this frequency may be higher than previously thought.

Evidently, current observational data are insufficient to completely rule out one or another formation channel, and indeed, multiple of them may be at work simultaneously. %In either the case of compact and main-sequence stars, 
Further studies of magnetic binary interactions can provide crucial insights to resolve this debate, of which binarity has shown to be a key element. % or "a key mystery of the debates"? 
%For example, the discovery of an abundance of doubly-magnetic systems would favour the fossil hypothesis, while ... may favour the merger scenario.

In particular, magnetism can play an important role in the dynamics of stellar, compact, and planetary systems, shaping the long-term evolution of their orbits \citep{Bourgoin_2022, Bromley2022}.
In gravitational wave (GW) astronomy, the upcoming generation of interferometric detectors such as the Laser Interferometer Space Antenna (LISA; \citep{LISA}) and the Einstein Telescope (ET; \citep{Maggiore2020}) will provide enough sensitivity to probe the interactions of compact systems and to characterise their magnetic attributes. On one hand, this can enable the composition of new magnetic population models and bring insight to the nature and origin of internal fields.
On the other hand, a careful understanding of the dynamics of magnetism in these systems is also required to avoid biases in the calibration of the instrument and in the interpretation of signals into physical parameters. Indeed, the secular (long-term) impact of magnetism on the orbits will manifest as a definitive signature on the GWs, which must be correctly accounted for \citep[][Savalle et al.\ in prep.]{Bourgoin_2022, Carvalho2022, Lira2022}.

It is thus imperative to study the secular evolution of the fields themselves, their binary coupling and the interplay with stellar orientation. In the case of stable, rigid fields, this translates to investigating the rotational dynamics of the stars, which may include their states of equilibrium.
In purely tidal-driven systems, spin motion and stability has long-since been determined by \cite{Hut_1980, Hut_1981}. In the case of star-planet systems, \cite{Damiani_2015} further explored the interaction between tides and magnetic braking --- where pressure-driven stellar winds give rise to the loss of angular momentum  --- and \cite{Strugarek_2017} determined the relative strengths of the tidal and magnetic effects in magnetic star-planet interactions.
The long-term effects of static dipole fields on stellar rotation, however, has yet to be completely explored.
%Amongst these, approximate analytical secular precession solutions were derived by \cite{Mikoczi_2021}, using an oblique dipole field model which considered the angle between the spins and orbital angular momentum constant. 
In this regard, the works of \cite{Pablo_2019} and \cite{King_1990} investigate the spin equilibrium under purely the dipolar interactions, but they neglect the interactions with orbital dynamics. In particular, they explore the cases where the obliquity between the dipole and spin axes is constrained to $0^\circ$ (aligned) and $90^\circ$ (perpendicular) respectively.

In this work, we direct our attention towards magnetic binary interactions --- in particular, towards stars with strong, stable, and predominantly dipolar fields. We consider the magnetic moments of these stars to be aligned with the stellar spin, and investigate the secular evolution of each star's orientation due to the mutual magnetic torques, through an effective orbit-averaged description. We provide criteria for determining whether magnetism dictates %\MCA{(when magnetism is said to dictate) evaluating the influence of magnetism on ?} 
the stellar rotational motion, which may then reflect on the secular evolution of the binary's orbits. Our study can be applied to any type of star system (MMS, WD, and NS constituents), as long as both components of the binary are magnetic and dominated by dipolar terms. In this way, it can be useful to gather further understanding of the formation processes of MMS stars as well as to ensure an efficient data processing of the LISA or ET observations in the context of compact-star binaries.

The paper is subdivided as follows. The complete physical setup and assumptions of our model are described in Sect.~\ref{sec:magnetic_binary_model}.
We derive, in Sect.~\ref{sec:secular_precession}, the effective secular orientation dynamics of the spins of each star due to dipole-dipole interactions. We then provide in Sect.~\ref{sec:equilibrium_states} an analysis of the equilibrium states and their respective stability, developing a simple analytical solution for spin precession which is valid for quasi-stable systems. Our results are verified numerically and applied to a system possibly satisfying our requirements (Sect.~\ref{sec:numerical_validation}). Finally, we compare the contrasting results with respect to the traditional instantaneous equilibrium (Sect.~\ref{sec:discussion}), highlighting each method's advantages and differences.

\paragraph{Notations and conventions.}
We presently introduce the notation used throughout the paper. For each vector $\vec u$ element of some vector space $U$, we represent its norm in light typeface $u = \norm{\vec u}$ and its direction by a hat $\uvec u = \vec u / u$. Whenever two vectors are parallel, we symbolize this relationship by $\smash{\phantom{\vec u~\perp~\vec v}}\mathllap{\vec u~\,\parallel\,~\vec v}$; and similarly, two perpendicular vectors are denoted by $\smash{\vec u~\perp~\vec v}$.
We represent the vector space dual to $U$ by a starred $U^*$. In this setting, we denote by an underscore $\underline{\vec u} \in U^*$ the associated canonical dot-product covector, that is, the linear functional from $U$ to $\R$ that satisfies $\underline{\vec u}~{:}~\vec v~{\mapsto}~(\vec u \cdot \vec v)$. For two vector spaces $U$ and $V$, we represent their Cartesian product by $U \times V = \{\, (\vec u, \vec v)\,;\, \vec u \in U,\, \vec v \in V \,\}$ and their tensor product by $U \otimes V = \{\, \vec u \otimes \vec v\,;\, \vec u \in U,\, \vec v \in V\,\}$. Finally, whenever two vector subspaces are disjoint $U \cap V = \{\, 0 \,\}$, their sum is direct and is denoted by $U \oplus V = \{\, \vec u + \vec v \mid \vec u \in U,\, \vec v \in V \,\}$. 
%\AB{Good idea the notations paragraph, but I will introduced it as a new section named `Notations and conventions' or just `notations' so that you finish the introduction with the description of the work you are about to present in the paper.}
%__________________________________________________________________

\section{Magnetic binary model}
\label{sec:magnetic_binary_model}
% \Chris{for GR formalism we may use points $\mathcal P$ and their coordinates $x(\mathcal P)$.}

In this section, we present the physical set up and the assumptions used throughout the paper. Then, we proceed to re-derive the instantaneous magnetically-driven precession equations that govern the rotational state of the system.

\FloatBarrier 
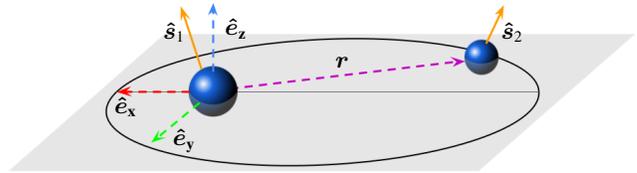
\begin{figure}[t]
\begin{center}
\end{center}
\setlength{\unitlength}{\linewidth}
\begin{picture}(1,0.3)(0,0)
 \put(0,0) {\includegraphics[width=1\linewidth]{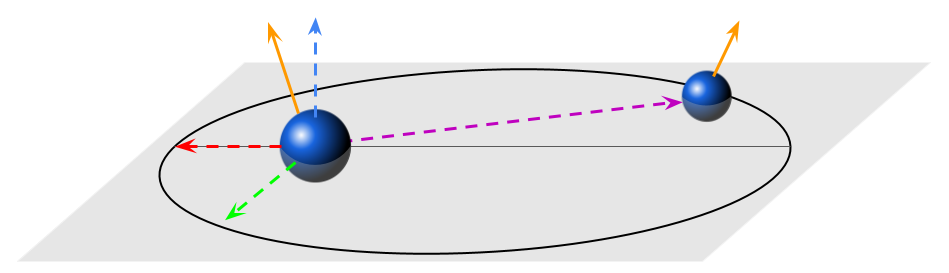}}
 % \linethickness{0.2mm}
 % \multiput(0,0)(.1,0){11}{\line(0,1){0.3}} 
 % \multiput(0,0)(0,.1){4}{\line(1,0){1}}
 % \linethickness{0.02mm}
 % \multiput(0,0)(.02,0){51}{\line(0,1){0.3}} 
 % \multiput(0,0)(0,.02){16}{\line(1,0){1}}
 \put(0.255,.215){\rotatebox{0}{$\uvec s_1$}}
 \put(0.775,.215){\rotatebox{0}{$\uvec s_2$}}
 \put(0.18,.10){\rotatebox{0}{$\basis{x}$}}
 \put(.27,.05){\rotatebox{0}{$\basis{y}$}}
 \put(.35,.22){\rotatebox{0}{$\basis{z}$}}
 \put(.52,.17){\rotatebox{0}{$\vec r$}} 
\end{picture}
\vspace{-0.2cm}
\caption{Binary system in the reference frame of the centre-of-mass. To simplify the drawing, the primary is placed at the origin (CM). We include the spin axis $\uvec s_\ell$ of each star ($\ell \in \{1,2\}$) and the basis $e_0 = (\basis x, \basis y, \basis z)$.}
\label{fig:binary_system_setup}
\end{figure}

Consider an isolated binary system of point-like magnetised bodies, dominated by non-relativistic motion. We assign to each body an index $\ell \in \{1, 2\}$, used throughout the paper, and we refer to them as the `primary' and `secondary', respectively. For each star we introduce a position $\vec x_\ell$, a mass $m_\ell$, a radius $R_\ell$, a magnetic field $\vec B_\ell$, and an intrinsic angular momentum (or spin) $\vec s_\ell$. 
We place ourselves in the reference frame of the centre-of-mass (CM) of the system, to which we attach a right-handed basis $e_0 = (\basis{x}, \basis{y}, \basis{z})$ spanning the Euclidean tangent space $E_3 \cong \R^3$. The elements of $e_0$ are chosen such that $\basis{x}$ points towards the direction of closest approach, $\basis{z}$ is orthogonal to the orbital plane, and $\basis{y}$ completes the basis. 
In this frame, the system can be viewed as an effective one-body problem, parametrised by the relative separation $\vec r = \vec x_2 - \vec x_1$.
In the absence of non-Keplerian perturbations, the CM frame will be inertial and the elements of $e_0$ will be static. The setup is illustrated in Fig.~\ref{fig:binary_system_setup}.

%Similarly, define a co-rotating basis $(\uvec{u}, \uvec{r}, \basis{z})$ such that the axis $\uvec{r}$ always points towards the secondary and the remaining orthonormal vector $\uvec u$ is in the plane of the orbit. 

We consider a stable magnetic field that is rigidly frozen into each star, compatible with general observations in massive stars and compact objects. The field is assumed to be predominantly dipolar, which captures most observed topologies (see the off-centered dipole model \citep{Borra_1982, Achilleos_1989}), although quadrupoles and octupoles have been detected in some cases \citep{Landstreet_2000, Wickramasinghe2000, Beuermann2007, Donati2009, Kochukhov2011, Landstreet2017}. In this scenario, we model $\vec B_\ell$ as a centered dipole, given in function of some point $\vec x$ outside the surface of the star:% with $\lvert \vec x - \vec x_\ell \rvert \gg R_\ell$
\begin{subequations}
\label{eq:B_magnetic_field}
\begin{align}
    &\vec B_1(\vec x) = \frac{\mu_0}{4\pi} \left( \frac{3\, \big(\vec \mu_1 \cdot (\vec x - \vec x_1) \big)\, (\vec x - \vec x_1)}{\lvert \vec x - \vec x_1 \rvert^5} - \frac{\vec \mu_1}{\lvert \vec x - \vec x_1 \rvert^3}\right) \text{,} \\
    &\vec B_2(\vec x) = \frac{\mu_0}{4\pi} \left( \frac{3\, \big(\vec \mu_2 \cdot (\vec x - \vec x_2) \big)\, (\vec x - \vec x_2)}{\lvert \vec x - \vec x_2 \rvert^5} - \frac{\vec \mu_2}{\lvert \vec x - \vec x_2 \rvert^3}\right) \text{,} 
\end{align}
\end{subequations}
where $\vec \mu_\ell$ is the magnetic dipole moment of each star, and $\mu_0$ is the vacuum permeability. 
The primary will feel the field $\vec B_2$ of its companion at relative position $\vec x = - \vec r$, and the secondary at $\vec x = \vec r$.
%The magnetic fields are linear expressions in function of the magnetic moments $\vec \mu_1$ and $\vec \mu_2$ respectively. 
It is convenient to express these fields in terms of the linear map $\mathcal B_{\vec r} : E_3 \to E_3$, which acts on the magnetic dipole moments according to:
\begin{align}
 \vec B_1(\vec r) = \frac{\mu_0}{4\pi} \mathcal B_{\vec r} (\vec \mu_1) \text{,} &&
 \vec B_2(- \vec r) = \frac{\mu_0}{4\pi} \mathcal B_{\vec r} (\vec \mu_2) \text{.}
\end{align}
$\mathcal B_{\vec r}$ can be identified with the symmetric tensor field with values in $E_3 \otimes E_3^*$, 
\begin{equation}
\mathcal B_{\vec r} = \frac{1}{r^3} (3\, \uvec{r} \otimes \raisebox{-0.2pt}{$\underline{\uvec{r}}$} - \mathbb I) \text{,}
\label{eq:B_operator}
\end{equation}
with $\mathbb I$ denoting the identity in $E_3$.

As is observed in the known doubly-magnetic MMS binary system $\epsilon$-Lupi \citep{Shultz_2015, Pablo_2019}, we examine the particular case where the fields are symmetric about the star's axis of rotation, with alignment between the magnetic dipole moment and the spin\footnote{Throughout the text, we shall assume $\mu_\ell$ positive, but $\mu_\ell < 0$ is also allowed, with appropriate sign changes in the equations.}:
\begin{equation}
    \vec \mu_\ell(t) = \mu_\ell \uvec s_\ell(t) \text{.}
\end{equation}
In practice, the magnetic moment amplitudes may be expressed in terms of the dipolar field evaluated at the poles $B_\ell^{\mathrm{p}} \coloneqq B_\ell(R_\ell \uvec s_\ell)$ [cf.\ Eq.~\eqref{eq:B_magnetic_field}], which can be observationally estimated via spectropolarimetry \citep[see e.g{.}][]{Shultz_2015}. We can thus invert Eq.~\eqref{eq:B_magnetic_field} to deduce:
\begin{equation}
    \mu_\ell = \frac{2 \pi}{\mu_0} B_\ell^{\mathrm{p}} R_\ell^3 \text.
    \label{eq:magnetic_moment}
\end{equation}

{
\renewcommand{\arraystretch}{1.35} 
\begin{table*}[tb]
    \caption{Typical binary system physical parameters (high-field range), and the corresponding dimensionless parameters: $\gamma^{\mathrm{fig}}~=~\Gamma^{\mathrm{fig}}/\Gamma^B$ the ratio between figure effects and magnetic torques [cf.\ Eq.~\eqref{eq:torque_fig_ratio}]; $\gamma^{\mathrm{tide}}~=~\Gamma^{\mathrm{tide}}/\Gamma^B$ the ratio between tidal and magnetic torques [cf.\ Eq.~\eqref{eq:torque_tide_ratio}]; $\eta = F^B/F^N$ the ratio between forces [Eq.~\eqref{eq:eta}]; and $\epsilon = P_{\mathrm{orb}}/\tau$ the ratio between dynamical timescales [Eqs.~\eqref{eq:timescale_torque}--\eqref{eq:timescale_ratio}]. We assume a binary system with two components of the same kind --- that is, MMS-MMS, WD-WD or NS-NS pairs. }       \label{tab:binary_parameters_orders_of_magnitude} 
    \centering
    \begin{tabular}{c|c|c|c|c|c|c|c|c|c}
    \hline\hline
         & \multicolumn{5}{c|}{Physical Parameters} &  \multicolumn{4}{c}{Ratios} \\ \hline
         Scenario & $m_1$, $m_2$ & $R_1$, $R_2$ & $\mathrlap{B_1}$\raisebox{2pt}{$\phantom{B}^p$}, $\mathrlap{B_2}$\raisebox{2pt}{$\phantom{B}^p$} & $P_1$, $P_2$ & $P_{\mathrm{orb}}$ \ghost{& $a$ [km]} & $\gamma^{\mathrm{fig}} = \Gamma^{\mathrm{fig}} / \Gamma^B$ & $\gamma^{\mathrm{tide}} = \Gamma^{\mathrm{tide}} / \Gamma^B$ & $\eta = F^B / F^N$ & $\epsilon = P_{\mathrm{orb}}/\tau$ \\\hline
         MMS & $10 \,M_\odot$ & $4.5 \,R_\odot$ & $10^4$\, G & $5$ d & $5$ d \ghost{& $2 \times 10^{10}$} & $2 \times 10^{8} \,J_2^\ell$ & $2 \times 10^{6\phantom{+}} \,k_2^\ell / Q_\ell$ & $5 \times 10^{-10}$ & $2 \times 10^{-9}$ \\
         WD &  $\phantom{0}1 \,M_\odot$ & $10^4$ km & $10^9$\, G & $1$ h & $10$ h \ghost{& $2 \times 10^{9\phantom{0}}$} & $2 \times 10^{6} \,J_2^\ell$ & $7 \times 10^{-1} \,k_2^\ell / Q_\ell$ & $1 \times 10^{-10}$ & $1 \times 10^{-7}$ \\
         NS & $1.4 \,M_\odot$ & $15$ km & $10^{15}$ G & $10$ min & $1$ h \ghost{& $5 \times 10^{8\phantom{0}}$} & $6 \times 10^{5} \,J_2^\ell$ & $7 \times 10^{-8} \,k_2^\ell / Q_\ell$ & $7 \times 10^{-15}$ & $5 \times 10^{-7}$ \\
         \hline
    \end{tabular}
\end{table*}
}
The magnetic field of each star will interact with the dipole of the companion, inducing the following torque: %\Chris{Possibility of introducing (1) the Lagrangian to "justify" the different types of interactions, or just (2) citing the sources. We may also (3) introduce the magnetic interaction energy at this point.}
\begin{subequations}
\label{eq:magnetic_torque}
\begin{align}
\vec\Gamma_1^{\mathrm{B}} &= \vec \mu_1 \times \vec B_2(- \vec r) \text{,} \\
\vec\Gamma_2^{\mathrm{B}} &= \vec \mu_2 \times \vec B_1(\vec r) \text{.}
\end{align}
\end{subequations}
Simultaneous contributions due to gravity exist. We introduce the total torque felt by $\ell$,
\begin{equation}
    \vec\Gamma_\ell = \vec\Gamma_\ell^{\mathrm{B}} +  \vec\Gamma_\ell^{\mathrm{fig}} +  \vec\Gamma_\ell^{\mathrm{tide}} + (\ldots) \text,
\end{equation}
where $\vec\Gamma_\ell^{\mathrm{B}}$, $\vec\Gamma_\ell^{\mathrm{fig}}$ and $\vec\Gamma_\ell^{\mathrm{tide}}$ are the respective contributions from the magnetic interaction, figure effects (rigid extended-body interactions), and tides. In order to  quantify the relative strength of the dipole-dipole interaction, we introduce the dimensionless parameter $\gamma$:
\begin{equation}
    \label{eq:torque_ratio}
    \gamma_\ell = \frac{\left\lvert \vec\Gamma_\ell - \vec\Gamma_\ell^{\mathrm{B}} \right\rvert}{\Gamma_\ell^{\mathrm{B}}} \leq \gamma_\ell^{\mathrm{fig}} + \gamma_\ell^{\mathrm{tide}} + (\ldots) \text,
\end{equation}
where $\gamma_\ell^{\mathrm{fig}} =  \Gamma_\ell^{\mathrm{fig}} / \Gamma_\ell^{\mathrm{B}}$ and $\gamma_\ell^{\mathrm{tide}} =  \Gamma_\ell^{\mathrm{tide}} / \Gamma_\ell^{\mathrm{B}}$ are the contributions to $\gamma$ due to figure effects and due to tides, respectively.
Our main interest lies in isolating the equilibrium dynamics of magnetic effects, and we shall therefore consider the regime where $\vec\Gamma_\ell^{\mathrm{B}}$ is dominant. More explicitly, we assume $\gamma_\ell^{\mathrm{fig}} \ll 1$ and $\gamma_\ell^{\mathrm{tide}} \ll 1$, for which we shall presently derive criteria.

The first assumption concerns the strength of figure effects. For a deformed extended body, classical gravitational torques up to quadrupole order have magnitudes roughly around $\Gamma_\ell^{\mathrm{fig}} \sim (3/2) \left({G m_1 m_2}/{r}\right) J_2^\ell \left( R_\ell/a\right)^2$, where $J_2^\ell$ is the dimensionless quadrupole moment, $a$ is the semi-major axis of the orbit, and $G$ is the gravitational constant \citep[see e.g{.}][]{poisson_will_gravity}. The corresponding contribution to $\gamma$ is 
\begin{equation}
    \label{eq:torque_fig_ratio}
    \gamma_\ell^{\mathrm{fig}} = \frac{\Gamma_\ell^{\mathrm{fig}}}{\Gamma_\ell^{\mathrm{B}}} \sim \frac{3 G \mu_0}{2 \pi}\frac{m_1 m_2}{B_1^{\mathrm{p}} B_2^{\mathrm{p}} R_1^3 R_2^3} R_\ell^2 J_2^\ell \text.
\end{equation}
Eq~\eqref{eq:torque_fig_ratio} shows that the ratio $\gamma_\ell^{\mathrm{fig}}$ is mainly scaled by the surface magnetic field strength, sphericity, and mean density of each stellar component.
% \Chris{The ideal scenario would be to define
% $$ \gamma_\ell = 1 - \frac{\Gamma^B}{\Gamma^{tot}}
% $$
% where $\Gamma^{tot}$ includes all torques acting on the system. When $\gamma \sim 0$ then magnetism dominates; when $\gamma \sim 1$ then magnetism is not important. With this definition, then the equation above is a particular case with $\Gamma^{tot} = \Gamma^B + \Gamma^N$
% }
In this work, we shall be considering perfectly spherical stars with $J_2^\ell = 0$, in which case we formally have no figure torques. In practice, the cutoff to $J_2^\ell$ below which rotation is driven by magnetism is given by
\begin{equation}
        J_2^\ell \lesssim \frac{2 \pi}{3 G \mu_0}\frac{B_1^{\mathrm{p}} B_2^{\mathrm{p}} R_1^3 R_2^3}{m_1 m_2} \frac{1}{R_\ell^2} \text,
        \label{eq:dimensionless_quad}
\end{equation}
and can be used as a criteria for the domain of validity of our models.

Similarly, we turn our attention to tidal interactions, which generate torques that scale with $\Gamma_\ell^{\mathrm{tide}}~{\sim}~6 ( G m_\mathfrak{m}^2 R_\ell^5 / a^6) (k_2^\ell / Q_\ell)$, where $m_\mathfrak{m}$ is the mass of the tide-inflicting body, $k_2^\ell$ is the gravitational Love number of $\ell$, and $Q_\ell$ is the tidal dissipation quality factor \citep[see e.g{.}][]{poisson_will_gravity, Strugarek_2017}. Then,
\begin{equation}
    \label{eq:torque_tide_ratio}
    \gamma_\ell^{\mathrm{tide}} = \frac{\Gamma_\ell^{\mathrm{tide}}}{\Gamma_\ell^{\mathrm{B}}} \sim \frac{6 G \mu_0}{\pi}\frac{m_\mathfrak{m}^2}{B_1^{\mathrm{p}} B_2^{\mathrm{p}} R_1^3 R_2^3} \frac{R_\ell^5}{a^3} \frac{k_2^\ell}{Q_\ell} \text,
\end{equation}
and the corresponding cutoff criteria for $k_2^\ell / Q_\ell$ is:
\begin{equation}
    \frac{k_2^\ell}{Q_\ell} \lesssim  \frac{\pi}{6 G \mu_0}\frac{B_1^{\mathrm{p}} B_2^{\mathrm{p}} R_1^3 R_2^3}{m_\mathfrak{m}^2} \frac{a^3}{R_\ell^5} \text.
    \label{eq:love_number}
\end{equation}

Table~\ref{tab:binary_parameters_orders_of_magnitude} shows typical values for $\gamma_\ell^{\mathrm{fig}}$ and $\gamma_\ell^{\mathrm{tide}}$ in MMS, WD and NS systems. At such distance scales ($a \sim 10^8$--$10^9$ km), the thresholds for $k_2 / Q$ are well within the range to allow magnetically-driven NS-NS systems, while tidal effects may have dominant contributions in MMS-MMS binaries, and WD-WDs lay somewhere in-between.
However, even for the most magnetic systems (NS-NS binaries), in order for the rotational dynamics not to be dominated by figure effects, we require that the quadrupole moment $J_2^\ell$ be at most on the order of $10^{-6}$.
%\AB{Could you comment a bit the ratio in the context of MMS, WD and NS from the values you derived in the Tab.?}

Regardless, under spherical, rigid star assumptions, the spins will suffer precession due to purely magnetic torques $\vec\Gamma_\ell^{\mathrm{B}}$; these torques are orthogonal to the intrinsic angular momentum, and hence the magnitude $s_\ell$ must be conserved. Substituting each term and normalizing by $s_\ell$ we obtain coordinate-free spin equations:
\begin{subequations}\label{eq:spin_precession_instantaneous}
\begin{align}
 \deriv{\uvec s_1}{t} &= - \alpha_1 \, \mathcal B_{\vec r} (\uvec s_2) \times \uvec s_1 \text{,}\\
 \deriv{\uvec s_2}{t} &= - \alpha_2 \, \mathcal B_{\vec r} (\uvec s_1) \times \uvec s_2 \text{,}
\end{align}
\end{subequations}
with $\alpha_\ell = \mu_0 \mu_1 \mu_2 / (4 \pi s_\ell)$. We denote Eq.~\eqref{eq:spin_precession_instantaneous} the `instantaneous' precession system. The spin axes are constrained to the unit sphere $\UnitSphere_2$% = \big\{ \vec u \in E_3\,;\, \norm{\vec u} = 1 \big\}$
, yielding a total of four degrees of freedom for the coupled system of equations. Each spin precesses around a (time-dependent) axis $\uvec \omega_\ell$ determined by the field direction $\uvec \omega_\ell \propto \mathcal B_r(\uvec s_\mathfrak{m})$ --- where $\mathfrak{m} \in \{1,2\}$, $\mathfrak{m} \neq \ell$ is the index of the companion star --- and have Larmor frequencies given by
% \begin{align}
%     \omega_1 = \norm{ \alpha_1\, \mathcal B_{\vec r} (\uvec s_2) } \text,
%     && \omega_2 = \norm{ \alpha_2 \mathcal B_{\vec r} (\uvec s_1) } \text,
% \end{align}
\begin{align}
    \omega_\ell = \norm{ \alpha_\ell\, \mathcal B_{\vec r} (\uvec s_\mathfrak{m}) } 
    %= \frac{\alpha_\ell}{r^3} \left(3 (\uvec r \cdot \uvec s_\mathfrak{m})^2 + 1 \right)^{1/2} 
    \sim \frac{\alpha_\ell}{r^3} \text.
    % && \omega_2 = \norm{ \alpha_2 \mathcal B_{\vec r} (\uvec s_1) } \text,
\end{align}

The orbital dynamics will induce periodic fluctuations on the separation $\vec r$, causing the precession axis and frequency to oscillate in time with period $P_{\mathrm{orb}}$. It is clear then that two distinct timescales will be involved in the dynamics of spin precession --- (1) a timescale corresponding to the orbital period $P_{\mathrm{orb}}$, manifesting in the `wobbles' of the axis $\uvec \omega_\ell$ and in the modulation of the frequency $\omega_\ell$; as well as (2) a timescale $\tau_\ell$ due to an average precession rate $\sec{\omega_\ell}$, which we define as:
\begin{equation}
\tau_\ell = \frac{2 \pi}{\sec{\omega_\ell}} \equiv \frac{2 \pi b^3}{\alpha_\ell} \,,
\label{eq:timescale_torque}
\end{equation}
where $b = \sqrt{r_{min} r_{max}}$ is the geometrical
%\footnote{The average precession rate $\sec{\omega_\ell} = \alpha_\ell \sec{1/r^{3}} \equiv \alpha_\ell / b^3$ has been defined with the \emph{geometrical} mean $b$ in order to maintain compatibility with the following sections. Indeed, with this definition, the characteristic timescale of the the effective orbit-averaged dynamics derived in Sect.~\ref{sec:secular_precession} can easily be seen to equal $\tau_\ell$. This particular choice corresponds therefore to the most suitable choice for a Keplerian orbit.}
mean between the separation at the pericenter and at the apocenter of the orbit. In an elliptical orbit, $b$ corresponds to the semi-minor axis.
%\Chris{in doubt between using this `more precise' definition or by instead considering $a$ instead of $b$ (similar orders of magnitude): $$\tau_\ell = \frac{2 \pi}{\omega_\ell} \sim \frac{2 \pi a^3}{\alpha_\ell}$$} 
% \AB{I would go on with the simplified version that captures the order of magnitude.} \Chris{I re-read the next section and it seems to get more confusing if I change this definition, it will affect a lot of things. I can't use orders of magnitude because the equations in the next section must be exact...}

\section{Secular precession}
\label{sec:secular_precession}

We are presently interested in determining the equilibrium configurations of the precession system and their stability. However, the instantaneous equilibrium obtained by equating \eqref{eq:spin_precession_instantaneous} to zero does not take into account the orbital dynamics; the strong dependence of the torques on the orbital position of each star implies that the configurations of equilibrium may largely fluctuate as the orbit evolves, which occurs in the fast timescale $P_{\mathrm{orb}}$. Indeed, a configuration that was momentarily stable at some point in the orbit may be disrupted by the orbital motion, leading to instability. Conversely, there may exist configurations where the spins oscillate in the fast timescale $P_{\mathrm{orb}}$, but on a longer timescale can be seen to be stable, due to an effective cancellation of the fluctuations.
It is therefore in our interest to search for states of secular (long-term) equilibrium and stability.
To do this we must eliminate the effects of these oscillatory terms of short period $P_{\mathrm{orb}}$, which can be performed by employing an orbital averaging scheme to obtain the effective dynamics. Variants of this method are widely adopted for determining secular solutions for the orbital motion \citep{Will_2017, Bourgoin_2022, Pound_2008, Gerosa_2015, poisson_will_gravity}.

For the binary systems considered in this work (magnetic MMS, WDs and NS), the two timescales ($P_{\mathrm{orb}}$ and $\tau_\ell$) of Eq.~\eqref{eq:spin_precession_instantaneous} will be distinct enough that their effects can be isolated.
Indeed, if the torque strength acting on a star is small enough, then its spin axis will not be significantly affected within a single orbital revolution. Intuitively, the impact of the torques on the spin axis is captured by the spin precession rate $\omega_\ell$. A larger precession rate (or smaller precession timescale $\tau_\ell$) implies faster changes to the axis $\uvec s_\ell$ due to stronger torques.
Thus, one may explicitly compute the characteristic time-ratio between the orbital timescale $P_{\mathrm{orb}}$ and the precession timescale $\tau_\ell$:
\begin{equation}
    \epsilon_\ell = \frac{P_{\mathrm{orb}}}{\tau_\ell} \sim  \frac{5 \pi}{2 G \mu_0} \frac{B_1^{\mathrm{p}} B_2^{\mathrm{p}} R_1^3 R_2^3}{(m_1 + m_2) m_\ell} \frac{1}{ R_\ell^2}\frac{P_\ell}{P_{\mathrm{orb}}} (1 - e^2)^{-3/2} \text,
    \label{eq:timescale_ratio}
\end{equation}
where $P_\ell$ is the rotational period of body $\ell$ and $e$ the eccentricity of the orbit. To derive the above order of magnitude relation, we have considered as a rough estimate that the mean separation is given by Kepler's third law $a^3 \sim G (m_1 + m_2) P_{\mathrm{orb}}^2 / (4\pi^2)$ and $b \sim a \sqrt{1 - e^2}$. We stress that these relationships are still valid as order-of-magnitude estimates for relativistic systems.
We have also estimated the spin magnitude for each star from that of a homogeneous sphere:
\begin{equation}
    s_\ell \sim \frac{4 \pi}{5} \frac{m_\ell R_\ell^2}{P_{\ell}} \text.
\end{equation}

In all three types of systems considered (Table~\ref{tab:binary_parameters_orders_of_magnitude}) we obtain very low values for $\epsilon$, namely $\epsilon_{\mathrm{MMS}} \sim 10^{-9}$, $\epsilon_{\mathrm{WD}} \sim 10^{-7}$ and $\epsilon_{\mathrm{NS}} \sim 10^{-6}$.
We therefore place ourselves in the scenario $\epsilon \ll 1$.
As discussed, in this scenario, the spin axes will suffer very little variations within the time-frame of a single orbit. We can therefore consider an effective precession dynamics which averages-out these small orbital oscillations.
Conversely, for systems with $\epsilon \gtrsim 1$ the two timescales cannot be decoupled in this manner. In this way, the time-ratio parameter $\epsilon$ can be used as a criteria for the validity of the averaging procedure that follows. 

For simplicity of the averaging model, we place ourselves in a classical Newtonian framework, although relativistic corrections are possible.
A rough estimate of the impact of magnetic fields on the orbit can be obtained by comparing the magnetic force $F^B \sim 3 \mu_0 \mu_1 \mu_2 / (4 \pi a^4)$ and the gravitational force $F^N \sim G m_1 m_2 / a^2$:
% \begin{equation}
%     \eta = \frac{F^B}{F^N} \sim \frac{3 \pi}{\mu_0} \left(\frac{1}{G}\right)^{5/3} \frac{B_1^{\mathrm{p}} B_2^{\mathrm{p}}
%     R_1^3 R_2^3}{m_1 m_2} \left(\frac{4 \pi^2}{(m_1 + m_2) P_{\mathrm{orb}}^2}\right)^{2/3}  \text.
%     \label{eq:eta}
% \end{equation}
\begin{equation}
    \eta = \frac{F^B}{F^N} \sim \frac{3 \pi}{G \mu_0} \frac{B_1^{\mathrm{p}} B_2^{\mathrm{p}}
    R_1^3 R_2^3}{m_1 m_2} \frac{1}{a^2}  \text.
    \label{eq:eta}
\end{equation}

Even for the most magnetic systems considered, this ratio is limited to $F^B / F^N \lesssim 10^{-10}$ (Table~\ref{tab:binary_parameters_orders_of_magnitude}).
We can therefore consider magnetism negligible in the orbital dynamics, both for MMS stars and compact systems. % or gravity dominant
In this setting, the orbital frame basis $(\basis{x}, \basis{y}, \basis{z})$ is inertial, the orbits are elliptical, and the separation $\vec r = r \uvec r$ can be parametrised as an ellipse in the centre of mass frame:
\begin{equation}
    \uvec r(f) = \basis{x} \cos{f} + \basis{y} \sin{f} \text{,} \qquad
    r(f) = \frac{a\, (1 - e^2)}{1 + e \sin{f}} \text{,}   
\label{eq:ellipse}
\end{equation}
where $f$ is the true anomaly, $a$ the semi-major axis and $e$ the eccentricity of the orbit.
The Keplerian solution determines the relationship between $f$ and $t$ (the time with respect to the reference pericenter passage):
\begin{subequations}\label{eq:Kepler_solution}
\begin{align}
    n t &\equiv E - e \sin E \mod 2\pi \text, \\
    E &= \arctantwo \left(e + \cos f, 
  \sqrt{1 - e^2} \sin f \right) \text,
\end{align}
\end{subequations}
where $\arctantwo$ represents the 2-argument inverse tangent, $E$ is the eccentric anomaly, and $n = \sqrt{{G (m_1 + m_2)}/{a^3}}$ is the mean angular motion.

To formalize our previous argument, consider the binary system at some given instant $t = t_0 + h$, with a short-timescale variation $h \in [0, P_{\mathrm{orb}}]$. In essence, we are considering that $h$ parametrises a single full orbital revolution of the binary, beginning at some time $t_0$. In this timeframe, the variations in the spin axis are bounded by
\begin{subequations}
\label{eq:secular_decomposition_smallness_of_variations}   
\begin{align}
    \lvert \uvec s_\ell(t_0 + h) -  \uvec s_\ell(t_0) \rvert 
    &= \left\lvert \int_0^h  \deriv{\uvec s_\ell}{t_0} (t_0 + u) \dd u \right\rvert \\
    &\leq \left(\, \sup_{u \in [0, h]} \left\lvert  \deriv{\uvec s_\ell}{t} (t_0 + u) \right\rvert \,\right) P_{\mathrm{orb}} \text{,}
\end{align} 
\end{subequations}
where the supremum of the derivative of $\uvec s_\ell$ can be obtained from the precession equation \eqref{eq:spin_precession_instantaneous}, with indices $\ell \in \{1, 2\}$ and $\mathfrak{m} \in \{1,2\}$, where $\mathfrak{m} \neq \ell$:
\begin{equation}
\left\lvert \deriv{\uvec s_\ell}{t} \right\rvert 
= \alpha_\ell \, \Big\lvert \mathcal B_{\vec r} (\uvec s_\mathfrak{m}) \times \uvec s_\ell \Big\rvert 
\leq \frac{1}{\tau_\ell} \left(\frac{1 + e}{1 - e}\right)^{3/2}
\text{.}
\end{equation}
It is easy to see [cf.\ Eq~\eqref{eq:B_operator}] that equality can be reached when $\uvec s_\mathfrak{m} = \uvec r$, $\uvec s_\ell \perp \uvec r$, and $r = r_{min} = 1/\big(a (1 - e)\big)$.

Eq.~\eqref{eq:spin_precession_instantaneous} may then be evaluated at time $t = t_0 + h$, and using the bounds obtained, expanded via $\uvec s_\ell(t_0 + h) = \uvec s_\ell(t_0) + O(\epsilon_\ell)$, from whence:
\begin{equation}
    \deriv{\uvec s_\ell}{t}(t_0 + h) = -  \alpha_\ell\, \mathcal B_{\vec r(t_0 + h)}\big(\uvec s_\mathfrak{m}(t_0)\big) \times\uvec s_\ell(t_0) + \frac{\alpha_\ell}{a^3 (1 - e^2)} O(\epsilon) \text,
\end{equation}
where we have explicitly included the temporal dependence of $\vec r = \vec r(t_0 + h)$ in the subscript of $\mathcal B_{\vec r}$, and defined $\epsilon = \max{(\epsilon_1, \epsilon_2)}$. Since the above equation is valid for any $h \in [0, P_{\mathrm{orb}}]$, integrating over a full orbit yields the secular spin equation
\begin{equation}
  \secderiv{\uvec s_\ell}{t}
  = - \alpha_\ell \sec{\mathcal B_{\vec r}} \big(\uvec s_\mathfrak{m} \big) \times \uvec s_\ell + \frac{\alpha_\ell}{a^3 (1 - e^2)} O(\epsilon) \text,
\end{equation}
where the orbital averaging operator is defined for some function of time $\xi$ as
\begin{equation}
    \langle \xi \rangle
    = \frac{1}{P_{\mathrm{orb}}} \int_0^{P_{\mathrlap{\mathrm{orb}}}} \xi(t_0 + h) \dd h 
    = \frac{n}{2\pi} \int_0^{2\mathrlap{\pi}} \tilde\xi(f_0 + f) \left(\deriv{t}{f}\right) \dd \mathrlap{f \text,}  % all this juggling to get the equation to fit.
\label{eq:secular_average}
\end{equation}
where $\tilde \xi(f(t)) = \xi(t)$ is the description of $\xi$ in terms of the true anomaly, and the Jacobian factor is found via implicit differentiation of Eq.~\eqref{eq:Kepler_solution}:
\begin{equation}
    \deriv{f}{t} = \frac{1}{n} \frac{(1 - e^2)^{3/2}}{(1 + e \cos f)^2} \text.
\end{equation}

In the Keplerian scenario the orbits are fixed, and the linear map $\mathcal B_{\vec r}$ --- which depends purely on the radial separation $\vec r$ --- is therefore $P_{\mathrm{orb}}$-periodic. Consequently, the orbital average $\sec{\mathcal B_{\vec r}}$ will be constant, independent of the secular time $t_0$. Plugging the expressions of $\mathcal B_{\vec r}$ [Eq.~\eqref{eq:B_operator}] and of the separation $\vec r$ [Eq.~\eqref{eq:ellipse}] into Eq.~\eqref{eq:secular_average}, we obtain the effective field tensor in the orbital frame basis $(\basis{x}, \basis{y}, \basis{z})$:
\begin{equation}
    \langle \mathcal B_{\vec r} \rangle
    = \frac{1}{2 a^3(1-e^2)^{3/2}} 
    % \begin{pmatrix}
    % 1 & 0 & 0 \\
    % 0 & 1 & 0 \\
    % 0 & 0 & -2
    % \end{pmatrix} \text{.}
    \Big( \mathbb I - 3\, \basis z \otimes \raisebox{-0.2pt}{$\underline{\basis z}$} \Big) 
    %\text{diag}(1,1,-2)
    \text{.}
    \label{eq:sec_B}
\end{equation}
Note that we have essentially averaged out the orbital oscillations of the magnetic field due to the short-timescale elliptical movement, obtaining a corresponding `average field' operator $\langle \mathcal B_{\vec r} \rangle$. As a consequence, the radial component of $\mathcal B_{\vec r}$ has been suppressed, leaving a dipolar effective field with a predominant component in the orthogonal direction $\basis z$, plus a weaker component aligned with the spin direction.

We denote the normalized term on the right-hand side of Eq.~\eqref{eq:sec_B} by $\bar{\mathcal B} = \mathbb I - 3\, \basis{z} \otimes \underline{\basis{z}}$. The final secular form of the spin precession equations is obtained by absorbing the constants together:
\begin{subequations}\label{eq:spin_spin_aligned_secular}
\begin{align}
  \secderiv{\uvec s_1}{t}
  &= - \nu_1\, \bar{\mathcal B} (\uvec s_2) \times \uvec s_1 \text, \\
  \secderiv{\uvec s_2}{t}
  &= - \nu_2\, \bar{\mathcal B} (\uvec s_1) \times \uvec s_2 \text,
\end{align}
\end{subequations}
where we have introduced the magnetic rotational frequencies $\nu_\ell$, coupling magnetism, spins and orbital parameters:
\begin{equation}
    \nu_\ell = \frac{\mu_0 \mu_1 \mu_2}{4 \pi} \frac{1}{s_\ell} \frac{1}{2 a^3(1-e^2)^{3/2}} \text{.}
\end{equation}

\begin{figure}
    \centering
    \includegraphics[width=\linewidth]{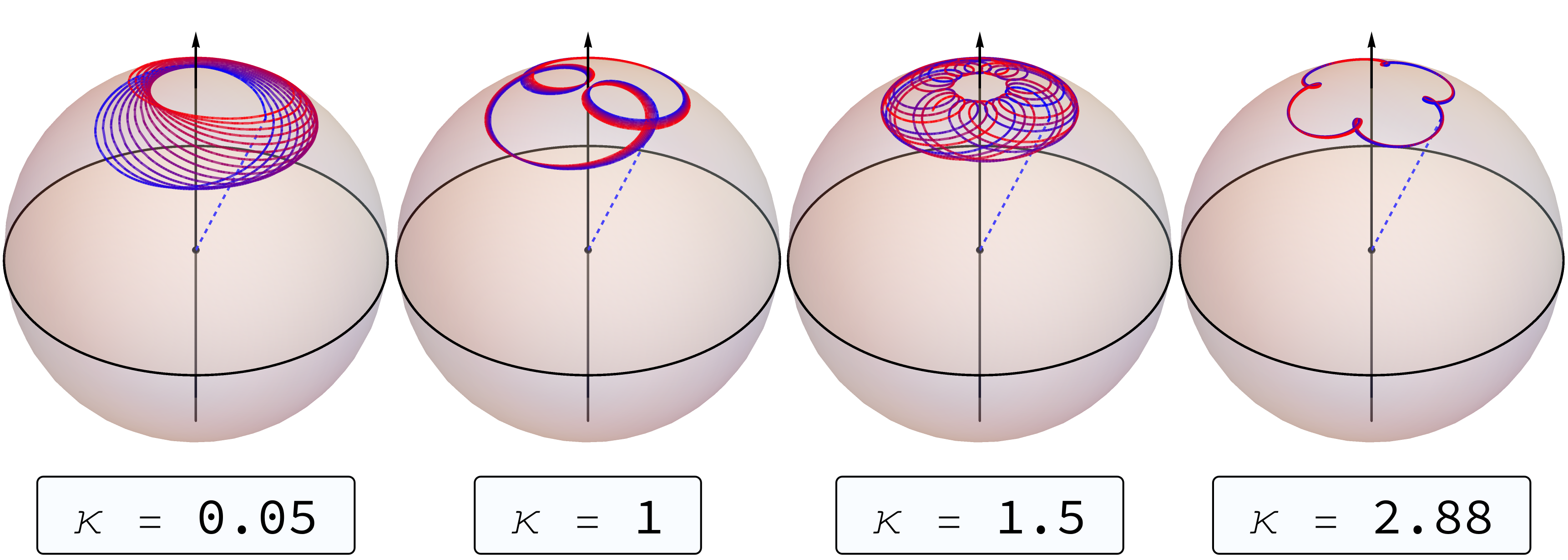}
    \caption{Sample trajectories for the secular evolution of the spin axis of the primary $\uvec s_1$, plotted against the unit sphere for different values of $\kappa = \nu_2/\nu_1$. The initial conditions are fixed, and shown for the primary as a dashed line from the origin to $\uvec s_1(0)$. The colours are interpolated between blue and red from initial time to $t_0 = 50\, \nu_1^{-1}$ respectively. The black axis represents the direction of $\basis z$. The behaviour for the secondary is analogous, up to a swap of initial conditions and $\kappa \mapsto \kappa^{-1} = \nu_1/\nu_2$.}
    \label{fig:example_spin_trajectories}
\end{figure}

The above system has four degrees of freedom, two for each spin, since the magnitudes of $\uvec s_\ell$ are conserved quantities of Eq.~\eqref{eq:spin_spin_aligned_secular}. Additionally, the magnetic interaction energy is also conserved, as will be discussed in Sect.~\ref{sec:sub:magnetic_interaction_energy}.
By performing a linear transformation on the secular time variable $t_0 \mapsto \nu_1^{-1} \,t_0$ it is possible to reduce the system to a single dimensionless parameter $\kappa = \nu_2 / \nu_1$, which depends only on the ratio between the two spin magnitudes. 
The parameter $\kappa$ will therefore completely control the dynamics of the system, producing a range of bounded trajectories such as illustrated in Fig.~\ref{fig:example_spin_trajectories}. These solutions are roughly epicyclic in nature, described by a predominant precessional motion around the axis $\basis z$ plus an important nutation component. When the respective frequencies of these two motions align as rational multiples of one another, the solutions become periodic. In the following section, we analyse the states of equilibrium of the secular dynamical system. We then proceed to analyse their stability and approximate solutions for trajectories similar to those presented in Fig.~\ref{fig:example_spin_trajectories}.

% \begin{figure}
%     \centering
%     \includegraphics[width=\linewidth]{img/example_spin_trajectories.png}
%     \setlength{\unitlength}{\linewidth}
%     \begin{picture}(0,0)(0,0)
%      \put(0.443,0.322){\rotatebox{0}{$\uvec s_1$}}
%      \put(0.388,0.0285){\rotatebox{0}{$\basis{x}$}}
%      \put(0.472,0.03){\rotatebox{0}{$\basis{y}$}}
%      \put(0.428,0.125){\rotatebox{0}{$\basis{z}$}}
%      \end{picture}
%     \caption{Sample trajectories for the secular evolution of the spin axis of the primary $\uvec s_1$, plotted against the unit sphere for different initial conditions and values of $\kappa$.
%     \todo{redo these figures with fixed initial conditions. The idea is to see the qualitative impact that kappa can have on the spin axis evolution.}}% The trajectories are conservative in terms of the average energy $\bar U_B$ and trace quasi-periodic “epicycle-like” motion. Equivalent conclusions are drawn for the secondary.}
%     \label{fig:example_spin_trajectories}
% \end{figure}

\section{Equilibrium states and stability}
\label{sec:equilibrium_states}

A `secular' equilibrium state at time $t_0$ is a pair of spins defined on an orbital period $(\uvec{ s_1 }, \uvec{ s_2 }) : [t_0, t_0 + P_{\mathrm{orb}}] \to E_3 \times E_3$ such that the average change in intrinsic angular momentum is zero:
\begin{align}
    \secderiv{\uvec s_1}{t}(t_0) = \secderiv{\uvec s_2}{t}(t_0) = 0 \text{.}
\end{align}

\begin{figure}
\centering
\includegraphics[width=\linewidth]{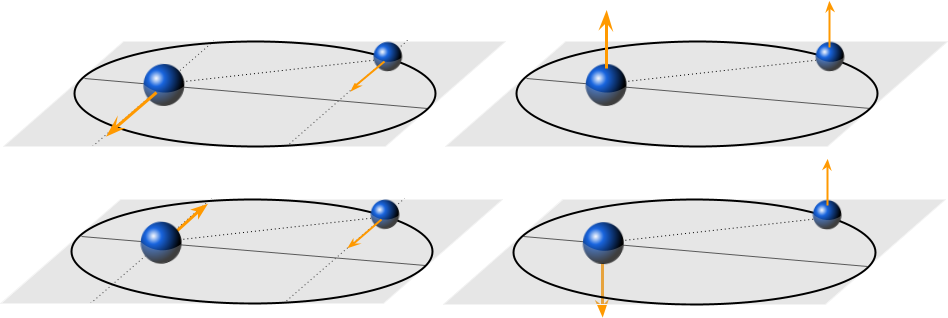}
\setlength{\unitlength}{\linewidth}
\begin{picture}(0,0)(0,0)
 \put(-0.295,0.172){\rotatebox{0}{$\uvec s_1$}}
 \put(-0.1,0.108){\rotatebox{0}{$\uvec s_2$}}
 \put(-0.232, 0.14){\rotatebox{0}{$\vec r$}}
 
 \put(-0.4,0.26){\rotatebox{0}{$\uvec s_1$}}
 \put(-0.095,0.275){\rotatebox{0}{$\uvec s_2$}}
 \put(-0.235, 0.305){\rotatebox{0}{$\vec r$}}
 
 \put(0.15,0.07){\rotatebox{0}{$\uvec s_1$}}
 \put(0.385,0.18){\rotatebox{0}{$\uvec s_2$}}
 \put(0.232, 0.14){\rotatebox{0}{$\vec r$}}
 
 \put(0.105,0.335){\rotatebox{0}{$\uvec s_1$}}
 \put(0.385,0.345){\rotatebox{0}{$\uvec s_2$}}
 \put(0.235, 0.305){\rotatebox{0}{$\vec r$}}
\end{picture}
\vspace{-0.2cm}
\caption{Equilibrium configurations of the spin axes. These correspond to directions that are either inside the orbital plane, in some arbitrary direction (case 1, left column), or orthogonal to the orbital plane (case 2, right column). In each case, the spins may either be parallel (top) or anti-parallel (bottom).
As discussed in Sect.~\ref{sec:equilibrium:stability}, only the anti-parallel orthogonal configuration  (bottom-right) is the only stable scenario.}
\label{fig:equilibrium_state_configs}
\end{figure}
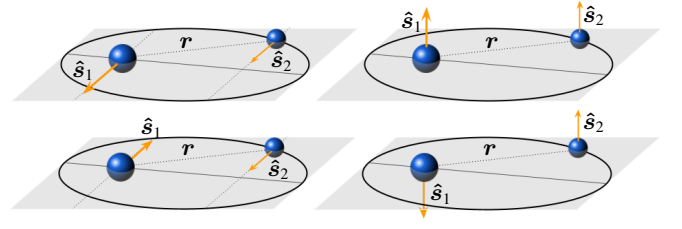

For a purely dipolar magnetic torque, it is clear from Eq.~\eqref{eq:spin_spin_aligned_secular} that this condition can only be achieved when the terms of each cross product are either parallel or zero, which we may write more concisely in the form:
\begin{align}
\bar{\mathcal B} (\uvec s_2) = \lambda_1 \uvec s_1 \text{,} &&
\bar{\mathcal B} (\uvec s_1) = \lambda_2 \uvec s_2 \text{.}
\end{align}
This corresponds to a singular-value problem, which can be solved with the explicit matrix form of $\bar{\mathcal B}$ in the orbital-frame basis. Two classes of solutions can be determined from the singular vectors of $\bar{\mathcal B}$, as described below.
In either class, the two spins must be parallel with each other, but they may point in the same direction or in reversed directions. Fig.~\ref{fig:equilibrium_state_configs} illustrates the full set of equilibrium configurations. % We define a parameter $\sigma = \uvec s_1 \cdot \uvec s_2 = \pm 1$ to describe this alignment.

\subsection*{Case 1. Equilibrium configurations in the orbital plane}

The first solution to the singular-value problem corresponds to an equilibrium configuration where the pair of spin axes are both contained inside the orbital plane:
\begin{align}
    \uvec s_1 = \sigma_1\, \uvec p \text{,} &&
    \uvec s_2 = \sigma_2\, \uvec p \text{,}
\label{eq:equilibrium_point_orbital}
\end{align}
for some unit vector $\uvec p$ in the plane of the orbit (i.e. $\uvec p \cdot \basis{z} = 0$). The two unitary parameters
$(\sigma_1, \sigma_2) \in \{-1, 1\} \times \{-1, 1\}$ describe the relative alignment between the two spins: they can either point in the same direction or in reversed directions. Notice that the set of all pairs $(\uvec s_1, \uvec s_2)$ which satisfy these conditions at a given moment in time forms a space of dimension 1.

\subsection*{Case 2. Equilibrium configurations orthogonal to the orbital plane}

The second solution to the singular-value problem corresponds to a configuration where both spins are orthogonal to the orbital plane --- in other words, parallel to the axis $\basis{z}$:
\begin{align}
    \uvec s_1 = \sigma_1 \, \basis{z} \text{,} &&
    \uvec s_2 = \sigma_2 \, \basis{z} \text{.}
\label{eq:equilibrium_point_orthogonal}
\end{align}
As in the previous case, the spins can be oriented in the same direction or in reverse directions according to the values of $\sigma_\ell$.

% \begin{figure*}
% \centering
% \includegraphics[width=\linewidth]{img/stability_configs.png}
% \setlength{\unitlength}{\linewidth}
% \begin{picture}(0,0)(0,0)
%  \put(-0.415,0.035){\rotatebox{0}{$\uvec s_1$}}
%  \put(-0.1,0.085){\rotatebox{0}{$\uvec s_2$}}
%  \put(-0.235, 0.11){\rotatebox{0}{$\uvec r$}}
 
%  \put(-0.335,0.35){\rotatebox{0}{$\uvec s_1$}}
%  \put(-0.095,0.36){\rotatebox{0}{$\uvec s_2$}}
%  \put(-0.235, 0.28){\rotatebox{0}{$\uvec r$}}
 
%  \put(0.17,0.14){\rotatebox{0}{$\uvec s_1$}}
%  \put(0.36,0.085){\rotatebox{0}{$\uvec s_2$}}
%  \put(0.235, 0.11){\rotatebox{0}{$\uvec r$}}
 
%  \put(0.127,0.165){\rotatebox{0}{$\uvec s_1$}}
%  \put(0.367,0.36){\rotatebox{0}{$\uvec s_2$}}
%  \put(0.235, 0.28){\rotatebox{0}{$\uvec r$}}
 
% \end{picture}
% \vspace{-0.2cm}
% \caption{Equilibrium configurations of the dipole axes. These correspond to aligned spin orientations orthogonal to the orbital plane (top) and contained within the orbital plane (bottom).
% Only the anti-parallel orthogonal configuration is stable (case 2, $\sigma = -1$, top-right).}
% \label{fig:equilibrium_state_configs}
% \end{figure*}

\subsection{Magnetic interaction energy between two dipoles}
\label{sec:sub:magnetic_interaction_energy}

The interaction energy between two magnetic dipoles is given by the symmetric expression below \citep[see e.g{.}][]{Pablo_2019}:
\begin{equation}
    U_{B}(\vec \mu_1, \vec \mu_2) = - \frac{\mu_0}{4 \pi} \vec \mu_1 \cdot \mathcal B_{\vec r} (\vec \mu_2) \text{.}
    \label{eq:magnetic_interaction_energy}
\end{equation} 
We shall denote $U_B$ the `instantaneous magnetic energy'. Conversely, one may take the orbital average of $U_B$, normalizing the resulting expression by a positive constant factor:
\begin{equation}
    \bar U_B(\uvec s_1, \uvec s_2) = - \uvec s_1 \cdot \bar{\mathcal B} (\uvec s_2) \text{.}
\label{eq:magnetic_interaction_energy_secular}
\end{equation}
We denote $\bar U_B$ the `secular magnetic energy'. As can be straightforwardly verified, $\bar U_B$ is a constant of motion of the secular system --- its time derivative vanishes for any pair $(\uvec s_1, \uvec s_2)$ of axes satisfying \eqref{eq:spin_spin_aligned_secular}.
Without considering additional forces or dissipation, the orbit-averaged precession system is conservative, and the motion is restricted to some level-curve of constant-energy $\bar U_B$.
In astrophysical systems, we expect dissipation due to radiation, tidal forces and internal frictions to bring the magnetic system to the lower energy states, by exchanging energy until eventually settling into a local minimum of $ \bar U_B$. As we shall see, this local minimum corresponds to a state of stability of the physical system.

In Appendix~\ref{appendix:optimization_U_B}, we recall that any symmetric bilinear form constrained to the unit $n$-sphere $U : \UnitSphere_n \times \UnitSphere_n \to \R$ is bounded by its largest-magnitude eigenvalue. Applying the principle to the secular magnetic energy [cf.\ Eq.~\eqref{eq:magnetic_interaction_energy_secular}], one obtains the bounds $-2 \leq \bar U_B \leq 2$.
The spin configurations in which these bounds are actually attained correspond to $\uvec s_\ell$ along the direction of the associated eigenvector --- that is, $\uvec s_\ell \parallel \basis z$. In fact, such configuration corresponds exactly to the equilibrium states orthogonal to the orbital plane as determined in the previous section [Eq.~\eqref{eq:equilibrium_point_orthogonal}]. In particular, the energy lower bounds are reached in the anti-parallel case ($\sigma_2 = - \sigma_1$), which as we shall see, is the most stable equilibrium point.
In the following section, we analyse the local stability of all the computed equilibrium states from the standpoint of the Hessian form of $\bar U_B$.

\subsection{Stability tests}
\label{sec:equilibrium:stability}

The stability of each equilibrium configuration can be analysed via the local convexity of the magnetic interaction energy. We evaluate the nature of each of the extrema --- minimum, maximum or saddle point --- by determining the sign of the Hessian at that point. We remind the reader of its definition.

Consider a real-valued function $f$ of $n$ real variables $\vec x = (x_1, \cdots, x_n)$ with all partial second-order derivatives. The Hessian of $f$ is a matrix-valued function $\mathcal H : \R^n \to \mathcal M_n(\R)$ defined as follows:
\begin{equation}
\mathcal H(\vec x) = \begin{pmatrix}
     \dfrac{\partial^2f}{\partial x_1^2}(\vec x) && \cdots && \dfrac{\partial^2f}{\partial x_1 \partial x_n}(\vec x) \\
     \vdots && \ddots && \vdots \\
     \dfrac{\partial^2f}{\partial x_n \partial x_1}(\vec x) && \cdots && \dfrac{\partial^2f}{\partial x_n^2}(\vec x) \\
    \end{pmatrix}.
\end{equation}
Evaluating the Hessian at some point $\vec x \in \R^n$ provides a description of the local convexity of $f$ at that point. 
If $\vec x^*$ is a critical point ($\nabla f(\vec x^*) = 0$) then $f$ can be locally approximated by a quadratic function
\begin{equation}
f(\vec x^* + \vec u) = f(\vec x^*) + \frac{1}{2} \vec u^T \mathcal H(\vec x^*)\, \vec u + O(u^3) \text.    
\end{equation}
% In particular, $H_{\vec x} \equiv H(\vec x)$ is positive-definite when
% \begin{equation}
% \vec u^T H_{\vec x} \, \vec u > 0 \,\quad \forall \vec y = (y_1, \ldots, y_n) \neq \vec 0\text,
% \end{equation} which implies $f$ is locally convex and $\vec x$ is an isolated point of local minimum.
% Conversely, $H_{\vec x}$ is negative-definite when
% \begin{equation}
% \vec u^T H_{\vec x} \, \vec u < \vec 0  \,,\quad  \forall \vec y = (y_1, \ldots, y_n)\neq \vec 0\text,    
% \end{equation}
% and hence $f$ is locally concave, with $\vec x$ is an isolated point of local maximum.
The Hessian matrix at critical point $\vec x^*$ can be decomposed into its eigenspace in order to obtain the principal directions of curvature. The sign of the corresponding eigenvalues determine whether each direction is stable or unstable.

For the problem at hand, we wish to study the critical points of the secular energy $\bar U_B$ as a function of spin direction.
For the following, we define the relative alignment between  spins $\uvec s_1$ and $\uvec s_2$:
\begin{equation}
    \sigma = \sigma_1 \sigma_2 \text.
\end{equation}

\subsubsection*{Case 1. Equilibrium configurations in the the orbital plane}

As we have seen, any pair of spin axes which are parallel to each other and contained within the orbital plane will correspond to a secular equilibrium state of the averaged system. 
Indeed, for any given $\uvec p \cdot \basis{z} = 0$ in Eq.~\eqref{eq:equilibrium_point_orbital} the energy takes the same value $\bar U_B(\sigma_1 \uvec p, \sigma_2 \uvec p) = \sigma$, depending only on the relative orientation of the two spins. The sign of $\sigma = \sigma_1 \sigma_2$ will be positive if the spins are facing the same direction or negative if they are in opposite directions. This invariance with respect to a choice of $\uvec p$ in fact reflects the more general axis-symmetry of the system with respect to rotations around the axis $\basis{z}$. Simultaneous rotations of both spins will leave the energy expression invariant.
In the case of this particular equilibrium configuration, the act of choosing one $\uvec p$ lying in the orbital plane over another $\uvec p'$ corresponds to rotating both spins simultaneously by the angle that takes $\uvec p' \mapsto \uvec p$. The set of all unit vectors $\uvec p$ in the orbital plane corresponds to a one-dimensional level-set of constant energy (a circle). To study the stability of this level-set we can reduce the degrees of freedom of the system from four (two for each spin axis, i.e.\ $\UnitSphere_2 \times \UnitSphere_2$) to three (by removing the rotational degree of freedom). In this reduced three-dimensional manifold, the circle becomes a point, and the remaining three directions of the tangent space determine the convexity of the energy.

As a first step, consider the spins parameterized in spherical coordinates,
\begin{equation}
    \uvec s_\ell = (\cos\psi_\ell \sin\theta_\ell, \sin\psi_\ell \sin\theta_\ell, \cos\theta_\ell)  \text{,} 
\end{equation}
with azimuth $\psi_\ell$ and angle $\theta_\ell$ measured from the north pole. 
In these coordinates, the secular energy takes the form
\begin{equation}
    \bar U_B = 2 \cos\theta_1 \cos\theta_2 - \cos(\psi_1 -\psi_2) \sin\theta_1 \sin\theta_2 \text{.}
\end{equation}
As discussed above, we can see clearly that simultaneous rotations in the azimuth $\psi_1$ and $\psi_2$ do not affect the energy. We can therefore look at the difference $\psi_1 - \psi_2 \coloneqq \Delta \psi$ and discard the redundant degree of freedom $\psi_1 + \psi_2$. At the equilibrium, the polar angles are $\theta_1 = \theta_2 = \pi/2$, and the azimuth is either $\Delta \psi = 0$ or $\Delta \psi = \pi$. We consider therefore the three directions of the tangent space of this reduced manifold. In this case, the Hessian matrix of the energy $\bar U_B$ with respect to the three variables $(\Delta \psi, \theta_1, \theta_2)$ has value:
\begin{equation}
    \mathcal H_1 = 
    \begin{pmatrix}
        \sigma & 0 & 0 \\
        0 & \sigma & 2 \\
        0 & 2 & \sigma \\
    \end{pmatrix}\text{.}
\end{equation}
The sign of $\sigma = \pm 1$ corresponds to the two cases of the spin directions being aligned or anti-aligned. For either sign, the Hessian has both positive and negative eigenvalues ($-1$, $1$ and $3 \sigma$). This implies a saddle point, which is unstable.

\subsubsection*{Case 2. Equilibrium configurations orthogonal to the orbital plane}

In the second case, we consider both spins to be orthogonal to the plane of the orbit. The spin axes $\uvec s_\ell$ therefore will be lying in the north or south pole of their corresponding unit spheres. A natural choice of coordinates for each spin in the neighbourhood of the poles is the Cartesian pair $x_\ell$, $y_\ell$, which with the unit-norm condition gives
\begin{equation}
    \uvec s_\ell = \left(x_\ell, y_\ell, \sigma_\ell \sqrt{1 - x_\ell^2 - y_\ell^2} \right) \text{.} 
\end{equation}
In these coordinates, the secular magnetic energy can be expressed in the form:
\begin{equation}
    \bar U_B = - x_1 x_2 - y_1 y_2 + 2 \sigma \sqrt{1 - x_1^2 - y_1^2} \sqrt{1 - x_2^2 - y_2^2} \text{.}
\end{equation}
The corresponding basis for the tangent space is the coordinate basis $\dd{x_\ell}$ and $\dd{y_\ell}$. With respect to these coordinates the Hessian of the energy takes the values:
\begin{equation}
    \mathcal H_2 =
    - \begin{pmatrix}
        2 \sigma & 0 & 1 & 0 \\
        0 & 2 \sigma & 0 & 1 \\
        1 & 0 & 2 \sigma & 0 \\
        0 & 1 & 0 & 2 \sigma \\
    \end{pmatrix} \text{.}
\end{equation}
The nature of the eigenvalues of the Hessian change according to the sign of $\sigma$: when the spins are aligned ($\sigma = + 1$), the Hessian is negative definite, with eigenvalues $\{ -3, -3, -1, -1 \}$, implying a point of maximum energy and therefore instability; when the spins are in opposing directions ($\sigma = - 1$), the eigenvalues are $\{ 1, 1, 3, 3 \}$ and the Hessian is positive definite, which implies a point of minimum energy and hence stability.

This suggests that given enough time and in the absence of stronger torques, magnetically interacting systems will naturally converge towards the stable anti-aligned orthogonal configuration. The timescale of this convergence will depend on the strength of the dissipation effects involved.

\begin{figure*}[!tb]
    \centering
    \includegraphics[width=\linewidth]{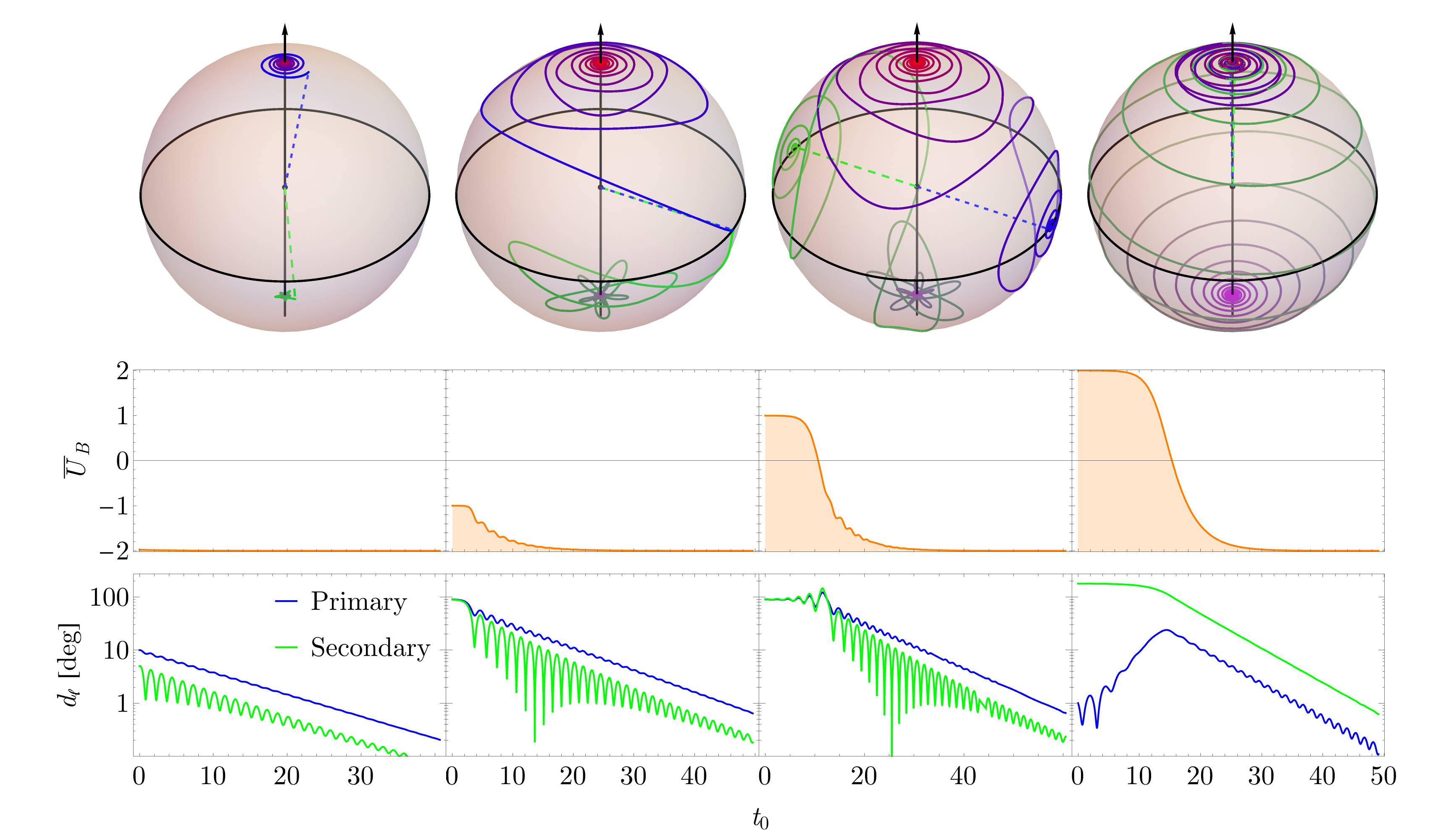}
    \caption{Evolution of the binary system in a secular timescale with dissipation introduced. Each column corresponds to a given initial condition from Table~\ref{tab:convergence_initial_conditions}, chosen in a neighbourhood of an equilibrium point. On the \emph{top} row: the trajectory of each spin axis $\uvec s_\ell$ is plotted against the unit sphere, in blue for the primary and green for the secondary. Colors are interpolated towards red from initial time to final time of convergence. Initial conditions for the spin axes are portrayed as dashed lines from the origin to $\uvec s_\ell(0)$. The orbital plane is represented in a darker shade with black contours, and the axis $\basis z$ as a vertical black arrow. On the \emph{middle} row: the secular magnetic interaction energy. On the \emph{bottom} row: the angular distance of each spin axis with respect to the stable equilibrium point $(\basis{z}, - \basis{z})$.}
\label{fig:convergence}
\end{figure*}

\subsection{Solutions near equilibrium}
\label{sec:solutions_near_equilibrium}

In Appendix~\ref{appendix:linearized_solutions}, we determine solutions near the stable equilibrium state. These solutions can be plugged in to derive complete secular orbital dynamics on systems which present strong magnetic torques, such as WD or NS binaries in the context of gravitational wave emission \citep[for further details, see e.g{.}][]{Bourgoin_2022, Lira2022}. 
We present below the main results:

The spins can be decomposed as $\uvec s_\ell = x_\ell\, \basis x + y_\ell\, \basis y + z_\ell\, \basis z$, where the orbital-plane components satisfy
\begin{subequations}
\begin{align}
    x_\ell(t) &= \rho_\ell^+ \cos{\Big( \omega_p^+ \,t + \phi_\ell^+ \Big)} + \rho_\ell^- \cos{\Big( \omega_p^- \,t + \phi_\ell^- \Big)} \text, \\
    y_\ell(t) &= \rho_\ell^+ \,\sin{\Big( \omega_p^+ \,t + \phi_\ell^+ \Big)} + \rho_\ell^- \,\sin{\Big( \omega_p^- \,t + \phi_\ell^- \Big)} \text,
\end{align}
\end{subequations}
and the orthogonal component satisfies
\begin{subequations}
\begin{align}
    z_1(t) = \frac{1}{2 \nu_2} \left(\zeta + \bar{{{Q}}}_z + \rho_z \cos{(\omega_z t + \phi_z)} \right) \text, \\
    z_2(t) = \frac{1}{2 \nu_1} \left( \zeta - \bar{{{Q}}}_z - \rho_z \cos{(\omega_z t + \phi_z)} \right) \text,
\end{align}
\end{subequations}
with real parameters $\rho_\ell^\pm$, $\phi_\ell^\pm$, $\rho_z$, $\phi_z$, dependent on initial conditions. The expressions of the constants $\zeta$ and $\bar{{{Q}}}_z$ and the frequencies $\omega_z$, $\omega_p^+$ and $\omega_p^-$ are given in the appendix.

\section{Numerical Validation}
\label{sec:numerical_validation}

We present in this section a numerical verification of the results that were obtained in Sect.~\ref{sec:equilibrium_states}.  In the first part, we demonstrate the derived (in-)stability of each equilibrium configuration. In the second part, we compare the obtained analytical solutions to numerical integration, in the neighbourhood of the stable equilibrium.

\subsection{Stability of the equilibrium configurations}

We artificially introduce dissipation into the dynamics of the secular system and numerically show that it is driven towards the stable states. Such a dissipative effect must preserve the norm condition on unit vectors and manifest as a friction when orientation changes. For this we include a time-delay term into the magnetic field that is felt by each companion star.
By our previous arguments in Sect.~\ref{sec:secular_precession}, it is direct to see that this term will propagate to the secular scale as follows:
\begin{subequations}
\label{eq:spin_spin_aligned_secular_dissipative}
\begin{align}
  \secderiv{\uvec s_1}{\tau}(\tau)
  &= \uvec{ s_1 }(\tau) \times \bar{\mathcal B} (\uvec s_2(\tau-\Delta \tau)) \text, \\
  \secderiv{\uvec s_2}{\tau}(\tau)
  &= \kappa\, \uvec{ s_2 }(\tau) \times \bar{\mathcal B} (\uvec s_1(\tau - \Delta \tau)) \text,
\end{align}
\end{subequations} 
where we have used the re-scaled dimensionless time $\tau \mapsto \nu_1^{-1} \tau$ and $\kappa = {\nu_2}/{\nu_1} = s_1 / s_2$ presented at the end of Sect.~\ref{sec:secular_precession}.
For convenience, we consider the delay to be an order of magnitude smaller than the average torque timescale, $\Delta \tau \sim 0.1 \,\kappa^{-1/2}$. This provides us with dissipative effects visible on the simulation timescales.

\begin{table}[b]
    \caption{Initial conditions in four distinct stability simulations (a--d). Each spin axis is parametrised in polar coordinates with azimuth $\psi_\ell$ and angle $\theta_\ell$ measured from the north pole.} \label{tab:convergence_initial_conditions}
    \centering
    \begin{tabular}{c|c|c|c|c}
        \hline\hline
        \null & $\theta_1$ ($^\circ$) & $\theta_2$ ($^\circ$) & $\phi_1$ ($^\circ$) & $\phi_2$ ($^\circ$) \\ \hline
        (a) & $10$ & $175$ & $0$ & $0$ \\
        (b) & $0$ & $1$ & $0$ & $0$ \\
        (c) & $90$ & $89$ & $0$ & $181$ \\
        (d) & $90$ & $91$ & $0$ & $1$ \\ \hline
    \end{tabular}
\end{table}

\begin{figure*}[!tb]
    \centering
    \includegraphics[width=1\linewidth]{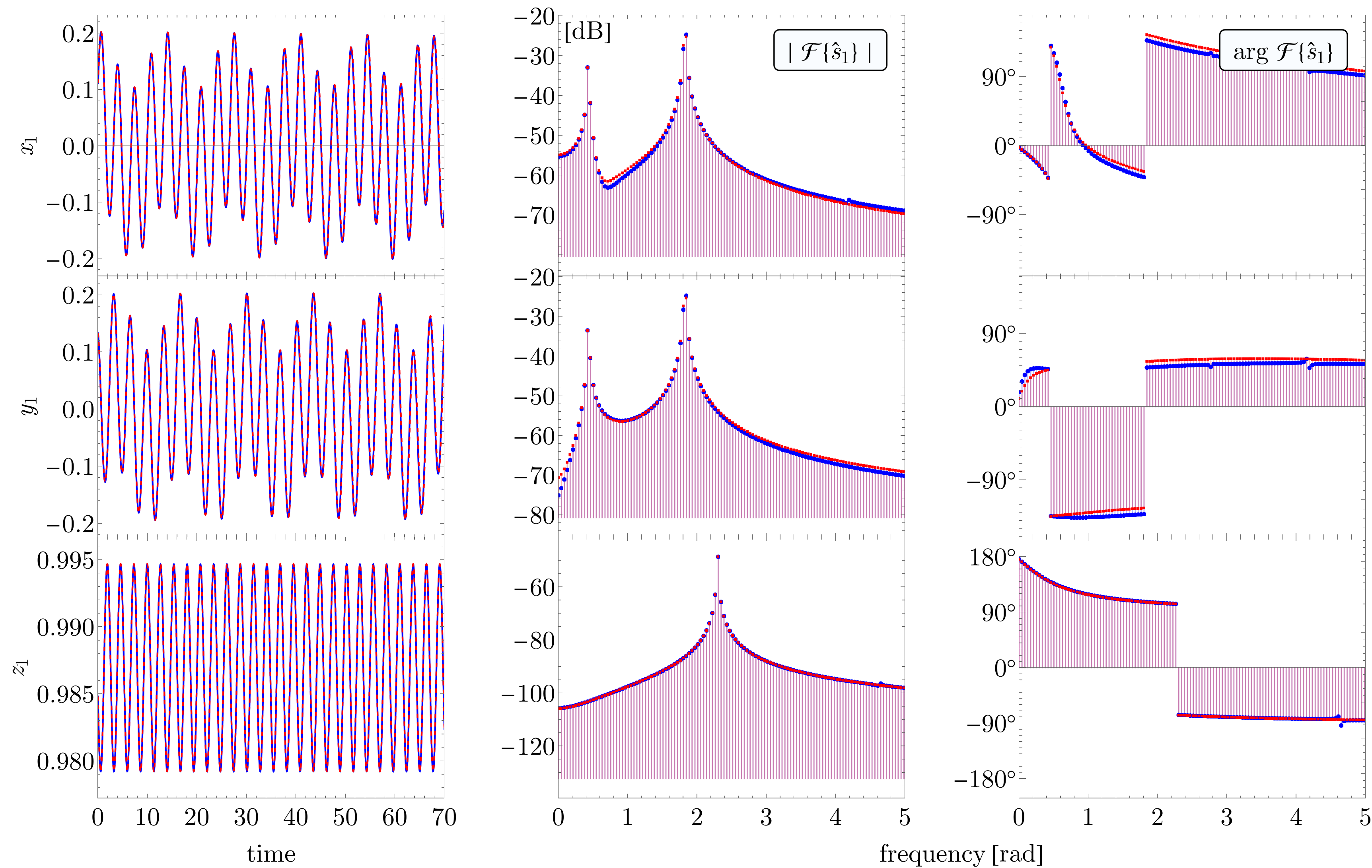}    
    \caption{Comparison between the analytical precession model (\emph{blue}) and numerical integration (\emph{red}), close to stable equilibrium. On the \emph{leftmost} column: the components of the spin of the primary, in the time domain. The corresponding Discrete Fourier Transform $\mathcal F \{ \uvec s_1 \}$ is given in absolute value (power spectrum in decibels, \emph{middle} column), and in complex argument (\emph{rightmost} column). The time and frequency axes are expressed in the re-scaled dimensionless units. }
    \label{fig:precession_analytical}
\end{figure*}

In order to assess stability, we consider two metrics: the secular magnetic interaction energy $\bar U_B$ [see Eq.~\eqref{eq:magnetic_interaction_energy_secular}]; and the angular distance to the stable equilibrium point $(\basis z, - \basis{z})$, which we define as:
\begin{subequations}
\begin{align}
    &d_{1} = \arg (\uvec s_1, + \basis{z}) \text, \\
    &d_{2} = \arg (\uvec s_2, - \basis{z}) \text,
\end{align}
\end{subequations}
where $\arg (\vec u, \vec v) = \arctantwo{ \big( \lvert \vec u \times \vec v \rvert , \vec u \cdot \vec v \big) }$ is the angle between any two vectors $\vec u$ and $\vec v$.

The dimensionless system is then integrated until convergence, for four sets of initial spin conditions (Table~\ref{tab:convergence_initial_conditions}), and with spin ratio $\kappa = 0.3$.
Each initial condition corresponds to an unstable equilibrium state plus a small perturbation on the order of ${\sim}1^\circ$. For completeness, we also include a perturbation of the stable state (${\sim}10^\circ$).
The resulting trajectories are plotted in Fig.~\ref{fig:convergence}, together with the time evolution of the secular energy $\bar U_B$ and of the angular distances $d_1$ and $d_2$. Even for a small initial angular perturbation, the spin axes diverge from their original unstable equilibrium and converge towards the stable, anti-aligned orthogonal configuration.

\subsection{Solutions near equilibrium}

In this part, we present a brief application of the analytical solutions of the secular equations that were exposed in Sect.~\ref{sec:solutions_near_equilibrium}. The equations are expressed in dimensionless time, and we adopt a spin ratio of $\kappa = 0.3$. The following initial conditions are considered (polar coordinates): inclinations $\theta_1 = 10^\circ$ and $\theta_2 = 172.5^\circ$ from the north pole, and azimuths $\psi_1 = 0$ and $\psi_2 = 50^\circ$. Figure~\ref{fig:precession_analytical} compares the obtained analytical expression of the primary to numerical integration, decomposed in the Cartesian basis $\uvec s_1(\tau) = x_1(\tau)\, \basis x + y_1(\tau)\, \basis y + z_1(\tau) \,\basis z$.
Matching peaks can be observed in the power spectra of both solutions, at the obtained angular frequencies (in dimensionless units) $\omega_p^- = 0.47$ rad and $\omega_p^+ = 1.86$ rad for the orbital components ($x_1, y_1$), and at $\omega_z = 2.34$ rad for the orthogonal component $z_1$ (cf. Appendix~\ref{appendix:linearized_solutions}).

\section{Discussion}
\label{sec:discussion}

We have so far defined the concept of a secular equilibrium state and applied it to obtain the equilibrium dynamics of binary systems with two magnetic components. In this section, we begin by discussing the fine points between secular and instantaneous equilibrium. Then, we apply our results to a real astrophysical scenario, the $\epsilon$-Lupi magnetic binary.

\subsection{Comparison between instantaneous and secular equilibrium states}
In Section~\ref{sec:equilibrium_states}, we defined the secular equilibrium as the spin configurations where the net torque over an orbital period is effectively zero. These states contrast with the \emph{instantaneous} equilibrium, the configurations of zero torque on the instantaneous precession system [Eq.~\eqref{eq:spin_precession_instantaneous}].
In the latter scenario, the spin dynamics are considered at a single moment in time and at a fixed orbital position. When orbital motion is introduced, an instantaneous equilibrium state may be destabilised. This can be seen in the following manner. Consider a bounded orbit parametrised by the osculating true anomaly $f = f(t)$. Expressing the spins in spherical coordinates, the instantaneous magnetic energy $U_B$ [Eq.~\eqref{eq:magnetic_interaction_energy}] takes the form:
\begin{equation}
U_B(t) = \frac{\mu_0 \mu_1 \mu_2}{4 \pi r^3} \Big(\uvec \mu_1 \cdot \uvec \mu_2 - 3 \sin\theta_1  \sin\theta_2 \cos(\psi_1 - f(t)) \cos(\psi_2 - f(t)) \Big) 
\end{equation}
Whereas the dot-product term on the right-hand side is invariant to an orbital translation of the bodies, the cosine terms will oscillate with the orbital dynamics.
Take two instants within the same orbital revolution, $t_1$ and $t_2$ such that the true anomaly at each instant equals $f(t_1) \equiv (\psi_1 + \psi_2)/2$ and $f(t_2) \equiv (\psi_1 + \psi_2 + \pi)/2$. Then the difference in energy between those two instants will be roughly
\begin{equation}
    U_B(t_2) - U_B(t_1) \sim \frac{3 \mu_0 \mu_1 \mu_2}{4 \pi a^3} \sin\theta_1 \sin\theta_2 \text,
\end{equation}
where we have substituted $r \sim a$.
We conclude that instantaneous equilibrium positions will indeed develop large energy oscillations in the timescale $t \sim P_{\mathrm{orb}}$, particularly when the polar angle $\theta_\ell$ is large (i.e. spin axes close to orbital plane alignment). For rapidly orbiting systems, this energy fluctuation occurs very quickly and destabilises the equilibrium of the point.

{
\renewcommand{\arraystretch}{1.25} 
\begin{table*}%[!b] 
    \caption{Comparison between the stability of each type of equilibrium state. The \emph{instantaneous} equilibrium states correspond to the spins either parallel or perpendicular to the orbital separation $\uvec r$. Analogously, the \emph{secular} states correspond to the spins either parallel or perpendicular to the axis $\basis z$. The energies $U_B$ and $\bar U_B$ are given in dimensionless units.
    }\label{tab:stability_comparison}
    \centering
    \begin{tabular}{ccccc|ccccc}
        \hline\hline
        \multicolumn{5}{c}{Instantaneous equilibria} & \multicolumn{5}{|c}{Secular equilibria} \\ \hline
        Config. & Cond. & $\sigma$ & $U_B$ & Stable & 
        Config. & Cond. & $\sigma$ & $\bar U_B$ & Stable \\ \hline
        
        $\rightrightarrows$ & $\uvec s_\ell \,\parallel\, \uvec r$ & $+1$ & $-2$ & yes &
        $\uparrow \uparrow$ & $\uvec s_\ell \,\parallel\, \basis z$ & $+1$ & $+2$ & no \\
        
        $\uparrow \downarrow$ & $\uvec s_\ell \perp \uvec r$ & $-1$ & $-1$ & no &
        $\leftrightarrows$ & $\uvec s_\ell \perp \basis z$ & $-1$ & $+1$ & no \\
        
        $\uparrow \uparrow$ & $\uvec s_\ell \perp \uvec r$ &  $+1$ & $+1$ & no &
        $\rightrightarrows$ & $\uvec s_\ell \perp \basis z$ & $+1$ & $-1$ & no \\
        
        $\leftrightarrows$ & $\uvec s_\ell \,\parallel\, \uvec r$ & $-1$ & $+2$ & no &
        $\uparrow \downarrow$ & $\uvec s_\ell \,\parallel\, \basis z$ & $-1$ & $-2$ & yes \\ \hline
    \end{tabular}
\end{table*}
}

There is a direct analogy between the states of secular equilibrium and those of instantaneous equilibrium. Recall the expression of the secular magnetic energy:
\begin{equation*}
    \bar U_B = -\, \uvec s_1 \cdot \bar{\mathcal B} (\uvec s_2) \text{,} \quad \bar{\mathcal B} = \mathbb I - 3\, \basis{z} \otimes \underline{\basis{z}} \text.
\end{equation*}
Such expression has been normalized by the positive constant $\mu_0 \mu_1 \mu_2 / 8 \pi b^3$ as discussed in Sect.~\ref{sec:equilibrium_states}.
We similarly normalize the expression of the instantaneous magnetic energy [cf. Eq.~\eqref{eq:magnetic_interaction_energy}] by the scalar $\mu_0 \mu_1 \mu_2 / 4 \pi r^3$ and obtain:
\begin{equation*}
U_B \propto +\, \uvec s_1 \cdot \tilde{\mathcal B}_{\vec r}(\uvec s_2) \text{,} \quad 
\tilde{\mathcal B}_{\vec r} = \mathbb I - 3\, \uvec{r} \otimes \raisebox{-0.2pt}{$\underline{\uvec r}$} \text{.}
\end{equation*}
Observe that the secular averaging procedure effectively produced a flip in the sign of the energy as well as a switch $\uvec r \leftrightarrow \basis z$. Analogously to how the secular equilibrium states of the binary are given by the singular vectors of $\bar {\mathcal B}$ (cf.\ Sect.~\ref{sec:equilibrium_states}), the instantaneous states are given by the singular vectors of $\tilde{\mathcal B}_{\vec r}$. Consequently, each state of instantaneous equilibrium has a secular counterpart. These correspond to spin axes aligned with the radial direction $\uvec r$ (resp.\ $\basis z$) or perpendicular to $\uvec r$ (resp.\ $\basis z$).
The stability of each state depends on the local convexity of the energy $U_B$ (resp.\ $\bar U_B$). We present in Table~\ref{tab:stability_comparison} a comparison between the obtained states for the two types of equilibrium.

\subsection{\texorpdfstring{$\epsilon$}{e}-Lupi}

We consider as a potential application case the $\epsilon$-Lupi inner binary system. $\epsilon$-Lupi is a ternary system composed of two close-range B-type companions Aa and Ab, plus a third distant companion dubbed $\epsilon$-Lupi B. Both stars of the $\epsilon$-Lupi A inner system are magnetic, making it the first and only currently known massive binary that has two magnetic components. Furthermore, the field of each star can be captured by a dipolar model with axes roughly parallel to the spins \citep{Shultz_2015}. The system also has a short orbital period of $P_{\mathrm{orb}} \sim 4.56 \text{ d}$, making it an excellent example to apply our model. \cite{Pablo_2019, Uytterhoeven_2005} obtained estimates for relevant stellar and orbital parameters, from which we adopt the values for the semi-major axis $a = 29.5 \,R_\odot$, eccentricity of $e = 0.28$, inclination $\iota = 21^\circ$, stellar masses of $m_1 = 9.0 \,M_\odot$ for the primary and $m_2 = 7.9 \,M_\odot$ for the secondary, and radii $R_1 = R_2 = 4.5 \,R_\odot$.

\cite{Shultz_2015} reported a dipolar field strength of at least $B_1^{\mathrm{p}} = 600 \,\text{G}$ and $B_2^{\mathrm{p}} = 900 \,\text{G}$ at the surface poles of the star, and projected rotational velocities at the equator $v_1 \sin i_1 = 37\text{ km s}^{-1}$ for the primary and $v_2 \sin i_2 = 27\text{ km s}^{-1}$ for the secondary.
By adopting the rotational inclination for each star equal to the orbital plane inclination $\iota \sim 21^\circ$, we obtain rotational periods on the order of $P_1 = 2.2 \,\text d$ and $P_2 = 3.0 \,\text d$.

From these physical parameters we may calculate the corresponding dimensionless ratios. The impact of figure effects can be assessed via $\gamma^{\mathrm{fig}}_\ell = 2 \times 10^{10} J_2^\ell$, whereas for tides $\gamma^{\mathrm{tide}}_1 = 2.5 \times 10^{8} \left( k_2^1 / Q_1 \right)$ and $\gamma^{\mathrm{tide}}_2 = 3.5 \times 10^{8} \left( k_2^2 / Q_2 \right)$. These expressions suggest that if $\epsilon$-Lupi is both highly symmetrical with $J_2^\ell \lesssim 6 \times 10^{-11}$, and has tidal parameters $k_2^\ell/Q_\ell \lesssim 3 \times 10^{-9}$, then the system's rotation is likely driven by magnetism. In this case, the smallness of $\eta = 5 \times 10^{-12}$ and $\epsilon = 4 \times 10^{-11}$ predict that the system will be driven towards the secular stable equilibrium in a timescale proportional to energy dissipation rates.
This outcome corresponds well to the expected state of $\epsilon$-Lupi based on observational data \citep{Pablo_2019}.

% \FloatBarrier
\section{Conclusions}

This paper has presented an analysis of the secular precession dynamics of binary systems under pure magnetic dipole-dipole interactions, considering an effective description with orbit-averaged motion. In particular, we have supposed spin-dipole alignment, perfect sphericity and tidal rigidity, and then derived criteria for assessing the validity of these assumptions, as well as the relative strengths of magnetic dipole interactions, tidal torques and figure effects. 
We have shown that this effective long-term description predicts a set of states of secular equilibrium which confront the traditional states of instantaneous magnetic equilibrium, where orbital dynamics are in fact neglected. Indeed, we have determined that there is a single secular state that is globally stable, corresponding to the configuration $\pm (\basis z, - \basis z)$ where the spin axes are reversed with respect to each other and orthogonal to the orbital plane. Conversely, the instantaneous state of radial alignment $\pm (\uvec r, \uvec r)$ is in fact only momentarily stable, since the orbital motion generates energy fluctuations and destabilizes the configuration. 
Our work can also be used to derive the long-term evolution of binary orbits providing an expected spin evolution in the absence of strong additional torques.

Our results hold for typical early-type MMS (such as the observed $\epsilon$-Lupi system), WD and NS binaries hosting dipolar fossil-like fields, where we expect long-term convergence towards the secularly stable state. 
Another interesting case of application is that of M dwarfs. For masses lower than $0.35 M_{\odot}$, M dwarfs are fully convective, unlike any other main-sequence stars, which renders the dynamo-driven topology unique \citep[e.g.][]{Dobleretal2006,Browning2008}. These low- and mid-mass M dwarfs are possible targets of application for our formalism since their magnetic fields often display intense dipolar components \citep[e.g.][]{Donatietal2008}, although intermittent higher-order multipolar components might also be present \citep[e.g.][and references therein]{Kochukhov2021}. Moreover, they have been observed forming binaries with two magnetic components \citep[e.g.][]{KochukhovLavail2017,KochukhovShulyak2019}. An interesting example of such binary systems is that of the YY Gem system \citep{KochukhovShulyak2019}, where each star has both a dipolar and multipolar components. The Zeeman-Doppler imaging analysis indeed revealed moderately complex global fields with a typical strength of 200-300 G, with dipolar components that are anti-aligned as predicted for fossil-type fields. These considerations hint that our work might be applied to any type of dipolar magnetic fields, either of fossil origin or triggered by a dynamo action, as long as its variation timescales are longer than the precession timescales. However, as mentioned in \cite{KochukhovShulyak2019}, the Zeeman intensification analysis suggests that the global fields of YY Gem may only comprise a few percent of their total magnetic fields, which highlights the need to consider multipoles in future studies. In addition, we point out that in typical M dwarf binary systems, gravitational interactions may control the orientation of the spins themselves. We can in fact compute (for a typical magnetic M dwarf: $m \sim 0.35 M_\odot$, $B \sim 1$ kG) the dimensionless parameters that measure the strength of extended-body gravitational interactions with respect to magnetic torques: $\gamma_{\mathrm{fig}} = 6 \times 10^{11} J_2$ and $\gamma_{\mathrm{tides}} = 7 \times 10^7 k_2/Q$ [Eqs.~(\ref{eq:torque_fig_ratio}, \ref{eq:torque_tide_ratio})]. Subsequently, we obtain the (quite tight) cutoffs $J_2 \leq 10^{-12}$ for the dimensionless quadrupole moment [Eq.~\eqref{eq:dimensionless_quad}], and $k_2 / Q \leq 10^{-8}$ for the tidal parameters [Eq.~\eqref{eq:love_number}], as the necessary parameters for magnetism to control the spin evolution.

Another key assumption from this work that warrants discussion is the alignment between the spin and magnetic axes of each star, since many misaligned systems are observed in nature \citep[e.g][]{Shultz2019}. In the case where the spin and magnetic dipole are anti-aligned [$\mu_\ell < 0$ in Eq.~\eqref{eq:magnetic_moment}], our general results still hold but spin directions must be flipped accordingly in the equations. More generally, this alignment constraint must be relaxed in future studies to explore the more general scenario of misaligned spin and magnetic axes. Assuming a rigid description where the magnetic axes rotate around the spins, a potentially direct case could be when the stellar rotation period $P_\ell$ is much shorter than the orbital period $P_{\mathrm{orb}}$. In such regime, the problem may be hierarchically split into three distinct timescales $P_\ell \ll P_{\mathrm{orb}} \ll \tau$, where $\tau$ is the `spin precession timescale' as described in Sect.~2 [Eq.~\eqref{eq:timescale_torque}]. The dynamics could then be formulated as an effective description when seen from the longer orbital timescale $P_{\mathrm{orb}}$, potentially reducing to the one explored in this work. Accordingly, one could expect our main results to still hold.

%where the system may be hierarchically developed into an effective description at the secular timescale $\tau$.} 
Further extensions of this work will also include abandoning the magnetostatic description and considering internal coupling of the fields with matter \citep[][]{Campbell2018}.
Finally, the precession dynamics and equilibrium states of the system may be investigated by directly taking into account not only magnetic forces but also competing figure effects and dynamical tides \citep[see example of such combined study in][]{Ahuiretal2021}. 

\begin{acknowledgements}
C.A. acknowledges the joint finantial support of Centre National d'Études Spatiales (CNES) and École Doctorale Astronomie et Astrophysique d'Ile de France (ED127 AAIF). This work was also supported by the Programme National GRAM, by PNPS (CNRS/INSU), by INP and IN2P3 co-funded by CNES, and by CNES LISA grants at CEA/IRFU. Authors are also grateful to S. Bouquillon for fruitful discussions. We are thankful to the anonymous reviewer for their constructive comments, which allowed us to improve our article.
\end{acknowledgements}

%\clearpage % For some reason there are huge empty floats and I need to flush them with clear page.

\bibliographystyle{aa} % style aa.bst
\bibliography{main} % main.bib

\begin{appendix}

\section{Optimization of bilinear forms}
\label{appendix:optimization_U_B}

In this appendix we recall the variational characterization of the singular value decomposition. Consider the bilinear form $U$ taking two unit vectors $U: \UnitSphere_n \times \UnitSphere_n \to \R$, represented under a basis $\big\{ \uvec e_1, ..., \uvec e_n \big\}$ via some matrix $U_{ij}$:
\begin{equation}
    U(\uvec u, \uvec v) = U_{ij} u^i v^j \text,    \label{eq:appendix_U_matrix_representation}
\end{equation}
where $\uvec u = u^i \uvec e_i$, $\uvec v = v^i \uvec e_i$, and sum over repeated indices is presupposed.

Since $U$ is continuous on the compact domain $\UnitSphere_n \times \UnitSphere_n$, it attains both a maximum and minimum value.
Indeed, consider the Lagrange function
 \begin{equation}
     g(\uvec u, \uvec v) = U(\uvec u, \uvec v) - \lambda_1 \big( \uvec u \cdot \uvec u - 1 \big) - \lambda_2 \big( \uvec v \cdot \uvec v - 1 \big) \text{,}
 \end{equation}
with multipliers $\lambda_1$ and $\lambda_2$.
The extrema $\uvec u_*$ and $\uvec v_*$ necessarily satisfy the stationary condition $\nabla g(\uvec u_*, \uvec v_*) = 0$,
 % \begin{equation}
 %     \dd g(\vec h_1, \vec h_2) = U(\vec h_1, \uvec v) - 2 \lambda_1 \,\uvec v \cdot \vec h_1 + U(\uvec u, \vec h_2) - 2 \lambda_2 \,\uvec u \cdot \vec h_2 \text{.}
 % \end{equation}
 which implies that for any tangent vectors $\vec h_1, \vec h_2$,
\begin{subequations}
\begin{align}
    &U(\vec h_1, \uvec v_*) = 2 \lambda_1 \,\uvec u_* \cdot \vec h_1 \text, \\
    &U(\uvec u_*, \vec h_2) = 2 \lambda_2 \,\uvec v_* \cdot \vec h_2 \text.    
\end{align}
\end{subequations}

From the matrix representation of $U$ [Eq.~\eqref{eq:appendix_U_matrix_representation}], we see that the solutions $\uvec u_*$ and $\uvec v_*$ correspond respectively to the left and right singular vectors of $U_{ij}$ --- that is, the unique set of vectors that satisfy
\begin{subequations}
\begin{align}
    U_{ij} u_*^i = \lambda v_*^i \text, &&U_{ij} v_*^j = \lambda u_*^j \text, 
\end{align}
\end{subequations}
where $\lambda = 2 \lambda_1 = 2 \lambda_2$ is the corresponding singular value of $U_{ij}$.

It is direct to see that the attained value $U_*~=~U(\uvec u_*, \uvec v_*)~=~\lambda$. From all singular-vector pairs $(\uvec u_*, \uvec v_*)$, the global maximum (or minimum) of $U$ is therefore reached by the candidate pair that has the highest (or lowest) corresponding singular value $\lambda$.
Note that whenever $U$ is symmetric, the singular vectors and eigenvectors of $U_{ij}$ coincide, and the singular values are equal to the magnitude of the eigenvalues.

\section{Linearized Solutions}
\label{appendix:linearized_solutions}

In this section, we formally derive secular solutions to the spin-spin equations,
\begin{subequations}
\label{eq:appendix:spin_spin}
\begin{align}
  \secderiv{\uvec s_1}{t}
  &= - \nu_1\, \bar{\mathcal B} (\uvec s_2) \times \uvec s_1 \text{,} \\
  \secderiv{\uvec s_2}{t}  
  &= - \nu_2\, \bar{\mathcal B} (\uvec s_1) \times \uvec s_2 \text{,}
\end{align}
\end{subequations}
and
\begin{equation*}
  \bar{\mathcal B} = \mathbb I - 3\, \basis z \otimes \raisebox{-0.2pt}{$\underline{\basis z}$} \text,
\end{equation*}
in a neighbourhood of the stable equilibrium point.
The solution obtained here is in fact more generally valid for any configuration close to the poles $(\uvec s_1, \uvec s_2) \approx (\sigma_1 \basis{z}, \sigma_2 \basis{z})$, including the unstable aligned orthogonal case (see Sect.~\ref{sec:equilibrium_states}).
As in the main text, we consider two indices $\ell, \mathfrak{m} \in \{1, 2\}$, with $\mathfrak{m} \neq \ell$, representing the pair of binaries permuted in some order.

In order to solve the system \eqref{eq:appendix:spin_spin}, we take advantage of the axial symmetry of the physical system around $\basis z$. We split the Euclidean vector space $E_3$ into an orbital plane $\Pi = \big\{x \,\basis x + y \,\basis y; (x, y) \in \R^2 \big\}$ plus a normal line $\Lambda = \big\{z \,\basis z; z \in \R \big\}$ such that $E_3 = \Pi \oplus \Lambda$. Rotations of the plane $\Pi$ leave \eqref{eq:appendix:spin_spin} invariant. 
% Projections of the spin axes onto each of these vector spaces may be computed via the decomposition
% \begin{align}
%     \uvec s_\ell = \text{proj}_\Pi(\uvec s_\ell) + \text{proj}_\Lambda(\uvec s_\ell) \text,
% \end{align}
% where
% \begin{align}
%     \text{proj}_\Pi = (\mathbb I - \basis z \otimes \raisebox{-0.2pt}{$\underline{\basis z}$}) \text, &&
%     \text{proj}_\Lambda = (\basis z \otimes \raisebox{-0.2pt}{$\underline{\basis z}$}) \text.
% \end{align}

We identify the line $\Lambda$ with the reals and the orbital plane $\Pi$ with the complex plane by introducing the linear isomorphism $\Phi : \Pi \times \Lambda \to \C \times \R$, which satisfies\footnote{The choice of real and imaginary axes is arbitrary. There is nothing special about $\basis x$ and $\basis y$. We could have equally taken any other pair of orthogonal axes in the orbital plane, and obtained equivalent results.}:
\begin{equation}
    \Phi(x \,\basis x + y \,\basis y + z \,\basis z) = (x + \im y, z) \text,
\end{equation}
where $\im$ is the imaginary number.
Eq.~\eqref{eq:appendix:spin_spin} can be expressed in the space $\C \times \R$ by declaring a new set of spin variables, which we define uniquely from $\Phi(\uvec s_\ell) = (\vec p_\ell, z_\ell)$. More explicitly, $\vec p_\ell$ corresponds to the orbital component of $\uvec s_\ell$, and $z_\ell$ corresponds to the projection of $\uvec s_\ell$ on the basis element $\basis z$:
\begin{align}
    \vec p_\ell &= (\uvec s_\ell \cdot \basis x) + \im\, (\uvec s_\ell \cdot \basis y) \text, \\
         z_\ell &= (\uvec s_\ell \cdot \basis z) \text.
\end{align}
This identification allows us to leverage the rotational symmetries of $\Pi$ through the algebraic structures of the complex numbers. In particular, through the isomorphism, vector dot- and cross-products in $E_3$ can be described in terms of complex multiplication and conjugation\footnote{For example, for two vectors contained in the orbital plane, $\vec u, \vec v \in \Pi$, the dot- and cross-product operations under the isomorphism are simply 
\begin{align*}
    \vec u \cdot \vec v &= \Real(\Phi_1^*(\vec u) \Phi_1(\vec v)) \text, \\
    \Phi_2(\vec u \times \vec v) &= \Imag(\Phi_1^*(\vec u) \Phi_1(\vec v)) \text.  
\end{align*}
with $\Phi = (\Phi_1, \Phi_2)$ and $\Phi_1^*$ the complex conjugate of $\Phi_1$.}.

For each body, we obtain the new and equivalent form of the spin precession equations:
\begin{align}
    &\deriv{\vec p_\ell}{t} = \im \, \nu_\ell \big( z_\ell\, \vec p_\mathfrak{m} + 2 z_\mathfrak{m}\, \vec p_\ell \big) \text, \label{eq:appendix:spin_spin_p}  \\
    &\deriv{z_\ell}{t} = \nu_\ell\, \Imag(\vec p_\ell^* \vec p_\mathfrak{m}) \text, \label{eq:appendix:spin_spin_z}
\end{align}
where 
$\Real, \Imag$ are the real and imaginary parts,
and the operation $\vec p \mapsto \vec p^*$ denotes complex conjugation. These equations are coupled by the four parameters $(\vec p_1, \vec p_2, z_1, z_2)$ with values in $\C^2 \times \R^2$, for a total dimensionality of 6. As discussed in Sect.~\ref{sec:secular_precession}, two degrees of freedom are redundant, constrained by the unit-norm conditions
\begin{equation}
    |\vec p_\ell|^2 + z_\ell^2 = 1 \text,
\end{equation}
and the magnetic interaction energy defines a conserved quantity:
\begin{equation}
    \bar U_B = 2 z_\ell z_\mathfrak{m} - \Real(\vec p_\ell^* \vec p_\mathfrak{m}) \text.
\end{equation}

The anti-symmetry of the equation \eqref{eq:appendix:spin_spin_z} with respect to a swap of indices $\ell \leftrightarrow m$ allows us to determine another first integral for the problem --- namely, the z-component of the total intrinsic angular momentum, $S_z = s_1 z_1 + s_2 z_2$. We introduce the quantity:
\begin{equation}
  \zeta = \nu_2\, z_1(t) + \nu_1\, z_2(t) = \frac{\mu_0 \mu_1 \mu_2}{8 \pi a^3(1-e^2)^{3/2}} \frac{1}{s_1 s_2} S_z \text.
\end{equation}
Since for any two complex numbers $\Imag(\vec u^*\vec v + \vec v^* \vec u) = 0$, it follows that the derivative of $\zeta$ vanishes.
We also introduce the anti-symmetric spin: 
\begin{equation}
    {{Q}}_z(t) = \nu_2 z_1(t) - \nu_1 z_2(t) \text,
\end{equation}
which is not a conserved quantity.

The spins of the two bodies can be decoupled in Eqs.~\eqref{eq:appendix:spin_spin_p} and \eqref{eq:appendix:spin_spin_z} via differentiation and substitution of the constraints. We obtain the second-order system of equations below, purely in terms of the variables $\vec p_1$, $\vec p_2$ and ${{Q}}_z$:
\begin{align}
    &\deriv{^2 {\vec p}_1 }{t^2}
        = \alpha_+ \deriv{\vec p_1}{t} + \beta_-\, \vec p_1
    \label{eq:appendix:p_1_Q_z} \text, \\
    &\deriv{^2 {\vec p}_2 }{t^2}
        = \alpha_- \deriv{\vec p_2}{t} + \beta_+\, \vec p_2
    \label{eq:appendix:p_2_Q_z} \text, \\
    &\deriv{^2 {{Q}}_z}{t^2}= a - b \,{{Q}}_z + 3 {{Q}}_z^3
    \label{eq:appendix:Q_z} \text,
\end{align}
with $\alpha_+$, $\alpha_-$, $\beta_+$, $\beta_-$ four functions defined via
\begin{align}
    &\alpha_\pm(t) = 2 \im \zeta\, \pm \frac{1}{\zeta + {{Q}}_z(t)}\deriv{{{Q}}_z}{t}(t) \text, \\
    &\beta_\pm(t) = \frac{3}{4}\left(\zeta^2 - {{Q}}_z^2(t) \right) \pm \frac{2 \im \zeta}{\zeta + {{Q}}_z(t)}\deriv{{{Q}}_z}{t}(t) \text,
\end{align}
and the constants $a \in \R$, $b \in \R^+$ given by
\begin{align}
    &a = (\nu_2^2 - \nu_1^2)\,\zeta \text, &&
    &b = \nu_1^2 + \nu_2^2 + 4 \bar U_B \nu_1 \nu_2 - \frac{3}{2} \zeta^2 \text.
\end{align}
% \begin{align}
%     &\deriv{^2\vec p_\ell}{t^2} =
%         \left(2 \im\, \zeta + \frac{\dot z_\ell}{z_\ell} \right)\, \deriv{\vec p_\ell}{t} + \left(3 \nu_1 \nu_2 \, z_\ell z_\mathfrak{m} - 2 \im\, \nu_\ell z_\mathfrak{m} \frac{\dot z_\ell}{z_\ell} + 2 \im \, \nu_\ell \dot z_\mathfrak{m} \right) \vec p_\ell \text, \label{eq:appendix:spin_2nd_order_p} \\
%     &\deriv{^2 z_\ell}{t^2} = 
%         {\zeta} \left(\nu_\mathfrak{m} + 2 \nu_\ell \bar U_B \right) - \left(\nu_\ell^2 + \nu_\mathfrak{m}^2 + 4 \bar U_B \nu_\ell \nu_\mathfrak{m} + 3 {\zeta}^2 \right) z_\ell \nonumber\\& + 9 \nu_\mathfrak{m} {\zeta} z_\ell^2 - 6 \nu_\mathfrak{m}^2 z_\ell^3 \text. \label{eq:appendix:spin_2nd_order_z}
% \end{align}
% which allows us to decouple the two spins.

To solve the full system of equations, we must first solve for ${{Q}}_z$ and then plug the obtained solution in the expression of $\alpha_\pm$ and $\beta_\pm$ in order to determine $\vec p_1$ and $\vec p_2$. In fact, there is an exact analytical solution for ${{Q}}_z$ in terms of elliptic functions. The resulting expression for ${{Q}}_z$ is somewhat lengthy, and subsequently solving \eqref{eq:appendix:p_1_Q_z}, \eqref{eq:appendix:p_2_Q_z} in this scenario proves to be challenging. Instead, we opt to present simpler and physically meaningful expressions for all the parameters, with domain of validity close to the poles.

Consider the expansion of Eqs.~\eqref{eq:appendix:p_1_Q_z} - \eqref{eq:appendix:Q_z} around some origin $\bar{{{Q}}}_z$, with ${{Q}}_z(t) = \bar{{{Q}}}_z + \delta {{Q}}_z(t)$. For a good choice of $\bar{{{Q}}}_z$, the resulting system may be truncated at low orders in $\delta {{Q}}_z$ to yield low-order solutions ${{Q}}_z^{(0)}$, ${{Q}}_z^{(1)}$, etc. In particular, we constrain $\bar{{{Q}}}_z$ to the domain of ${{Q}}_z$ by choosing $\bar{{{Q}}}_z = {{Q}}_z(t_0)$ for a reference time $t_0$. The error incurred from this expansion will depend on the largeness of the variations $\delta {{Q}}_z$ --- which will be considerably small in the neighbourhood of the poles. 
To understand this, consider a spin variation from $t_0$ to $t$, $\delta \uvec s_\ell(t) = \uvec s_\ell(t) - \uvec s_\ell(t_0)$.
Due to the geometry of the unit sphere, this variation will propagate along the z component to some $\lvert \delta z_\ell(t) \rvert \leq \lvert \delta \uvec s_\ell(t)\rvert \sin{\theta_\ell} \leq 2 \sin^2{\theta_\ell}$, where $\theta_\ell$ is the maximum attained polar inclination, that is, $\sin \theta_\ell = \sup_t |\uvec s_\ell(t) \cdot \basis z|$. For a solution close to the poles, we consider the polar angle $\theta_\ell$ a small parameter. In this case, the perturbation $\delta {{Q}}_z(t) = \nu_2 \delta z_1(t)  - \nu_1 \delta z_2(t)$ will also remain similarly bounded. With these considerations, we present below the solutions for low orders of $\delta {{Q}}$.

\subsection*{Orthogonal component}

As discussed, we begin by determining an expression for the decoupled variable ${{Q}}_z$ [Eq.~\eqref{eq:appendix:Q_z}]. 
Until this point arbitrary, we fix the choice of $\bar{{{Q}}}_z$ to best fit \eqref{eq:appendix:Q_z} and minimize $\delta {{Q}}_z$. A good expansion parameter $\bar{{{Q}}}_z$ is in fact the zero-order constant solution ${{Q}}_z^{(0)}$, the unique real value that satisfies the third-degree algebraic equation:
\begin{equation}
    a - b \, {{Q}}_z^{(0)} + 3 \left( {{Q}}_z^{(0)} \right)^3 = 0 \text,
\end{equation}
whereas the first-order solution is
\begin{equation}
    {{Q}}_z^{(1)}(t) = {{Q}}_z^{(0)} + \rho_z \cos{(\omega_z t + \phi_z)} \text,
\label{eq:analytical_spin_Q}
\end{equation}
with
\begin{subequations}
    \begin{align}
        &\omega_z = \left( b + \frac{9}{2} \bar{{{Q}}}_z^2 \right)^{1/2} \text, \\
        &\phi_z = \arctantwo{\left ({{Q}}_z(0)-\bar{{{Q}}}_z, - \frac{1}{\omega_z} \deriv{{{Q}}_z}{t}(0)\right)} \text, \\
        &\rho_z = \sqrt{\Big({{Q}}_z(0)-\bar{{{Q}}}_z \Big)^2 + \left(\frac{1}{\omega_z} \deriv{{{Q}}_z}{t}(0) \right)^2} \text.
        % & {{Q}}_z(0) = \left( \nu_2 \, \uvec s_1(0) - \nu_1 \uvec s_2(0) \right) \cdot \basis{z} \text, \\
        % &\dot {{Q}}_z(0) = 2 \nu_1 \nu_2 \, \uvec s_1(0) \times \uvec s_2(0) \cdot \basis{z} \text.
    \end{align}
\end{subequations}
We see \emph{a posteriori} that the choice of $\bar{{{Q}}}_z = {{Q}}_z^{(0)}$ is in fact the average of the first-order solution ${{Q}}_z^{(1)}$.

% \todo{Do some quick simulations on the solutions.}

\subsection*{Orbital component}

For the orbital part, we consider a zero-order approximation in $\delta {{Q}}$. In this scenario, each component $\vec p_\ell^{(0)}$ obeys a constant-coefficient linear ordinary differential equation, with
\begin{align}
    &\alpha_+^{(0)}(t) = \alpha_-^{(0)}(t) = 2 \im \zeta  \text, \\
    &\beta_+^{(0)}(t) = \beta_-^{(0)}(t) = \frac{3}{4}\left(\zeta^2 - \bar{{{Q}}}_z^2 \right) \text.
\end{align}

The corresponding orbital plane solutions are given by superposed complex rotations:
% \begin{equation}
%   \deriv{^2 {\vec p}_\ell }{t^2}
%   = 2 \im\, \zeta \, \dot {\vec p}_\ell + \left(3 \nu_1 \nu_2 \, \bar z_1 \bar z_2 \right) \vec p_\ell 
% \end{equation}
\begin{subequations}
\begin{align}
    &\vec p_1^{(0)}(t) = c_1^+ \, \e^{\im \, \omega_p^+ \,t} + c_1^- \, \e^{\im \omega_p^- \,t} \text,
    \label{eq:analytical_spin_p_1} \\
    &\vec p_2^{(0)}(t) = c_2^+ \e^{\im \, \omega_p^+ \,t} + c_2^- \e^{\im \omega_p^- \,t} \text,
    \label{eq:analytical_spin_p_2}
\end{align}    
\end{subequations}
with two oscillating frequencies
\begin{equation}
    \omega_p^\pm \coloneqq \zeta \pm \frac{1}{2} \sqrt{\zeta ^2 + 3 \bar{{{Q}}}_z^2 } \quad \in \mathbb R^+ \text,
\end{equation}
and four constants $c_1^+, c_1^-, c_2^+, c_2^-$ in the unit complex disk $D = \big\{ \mathrm z \in \C : |\mathrm z| \leq 1 \big\}$; they can be computed from the initial conditions as:
\begin{equation}
	c_\ell^\pm = \frac{\mp 1}{\omega_p^+ - \omega_p^-}\left(\im \deriv{\vec p_\ell}{t}(0) + \omega_p^\mp \vec p_\ell(0) \right) \text.
\end{equation}
% \begin{align}
%     c_\ell^+ = \frac{-1}{\omega_p^+ - \omega_p^-}\left(\im \deriv{\vec p_\ell}{t}(0) + \omega_p^- \vec p_\ell(0) \right) \text, \\
% 	c_\ell^- = \frac{ 1}{\omega_p^+ - \omega_p^-}\left(\im \deriv{\vec p_\ell}{t}(0) + \omega_p^+ \vec p_\ell(0) \right) \text.
% \end{align}
The calculations may naturally be extended to first-order solutions $\vec p_\ell^{(1)}$, which include a higher number of harmonic frequencies.

We now return to the original Euclidean space $E_3$. The spin can be broken down into the corresponding components of the basis $e_0$ by inverting the isomorphism $\Phi$:
\begin{equation*}
    \uvec s_\ell = \Phi^{-1}(\vec p_\ell, z_\ell) = \Real(\vec p_\ell)\, \basis x + \Imag(\vec p_\ell)\, \basis y + z_\ell\, \basis z \text.
\end{equation*}
From the above solutions, the component $z_\ell$ of the spins can then be retrieved from \eqref{eq:analytical_spin_Q} via the relations
\begin{equation}
    z_1^{(1)}(t) = \frac{1}{2 \nu_2} \left(\zeta + {{Q}}_z^{(1)}(t) \right) \text, \\
    z_2^{(1)}(t) = \frac{1}{2 \nu_1} \left( \zeta - {{Q}}_z^{(1)}(t) \right) \text, \\
    \label{eq:analytical_spin_z}
\end{equation}
and the two orbital components:
\begin{subequations}
\begin{align}
    \Real(\vec p_\ell(t)) &= \rho_\ell^+ \cos{\Big( \omega_p^+ \,t + \phi_\ell^+ \Big)} + \rho_\ell^- \cos{\Big( \omega_p^- \,t + \phi_\ell^- \Big)} \text, \\
    \Imag(\vec p_\ell(t)) &= \rho_\ell^+ \,\sin{\Big( \omega_p^+ \,t + \phi_\ell^+ \Big)} + \rho_\ell^- \,\sin{\Big( \omega_p^- \,t + \phi_\ell^- \Big)} \text,
\end{align}
\end{subequations}
with $\rho_\ell^\pm = |c_\ell^\pm| \in [0, 1]$ and $\phi_\ell^\pm = \arg c_\ell^\pm \in [0, 2\pi]$. 

\end{appendix}

%\vspace{1cm}
%\newpage

\end{document}